\documentclass[12pt]{article}
\usepackage{graphicx}
\usepackage{amsmath}
\usepackage{amssymb}
\usepackage{theorem}

\expandafter\let\csname equation*\endcsname\relax
\expandafter\let\csname endequation*\endcsname\relax
\usepackage{amsmath}
\usepackage{psfrag}

\usepackage{graphicx}
\usepackage{amssymb}
\usepackage{theorem}
\newtheorem{thm}{Theorem}

\newcommand{\E}{{\mathbb E}}

\title{On the detection of superdiffusive behaviour in time series}

\author{
Georg A. Gottwald\thanks{School of Mathematics and Statistics, University of Sydney, Sydney 2006 NSW, Australia, georg.gottwald@sydney.edu.au}
\and
Ian Melbourne\thanks{Mathematics Institute, University of Warwick, Coventry, CV4 7AL, UK, i.melbourne@warwick.ac.uk}
}

\date{19 November 2016}

\begin{document}

\maketitle

\begin{abstract}
We present a new method for detecting superdiffusive behaviour and for determining rates of superdiffusion in time series data. 
Our method applies equally to stochastic and deterministic time series data (with no prior knowledge required of the nature of the data) and relies on one realisation (ie one sample path) of the process. 
Linear drift effects are automatically removed without any preprocessing. We show numerical results for time series constructed from {\em{i.i.d.}}\ $\alpha$-stable random variables and from deterministic weakly chaotic maps. We compare our method with the standard method of estimating the growth rate of the mean-square displacement as well as the $p$-variation method, maximum likelihood, quantile matching and linear regression of the empirical characteristic function.
\end{abstract}


\section{Introduction}
\noindent
The ubiquity of normal diffusion can be understood by the Central Limit Theorem which, roughly speaking, states that an appropriately scaled sum of many independent identically distributed random variables with finite variance converges in distribution to a normally distributed random variable. Hence, the erratic motion of a grain of pollen suspended in water, as observed by the botanist Brown in 1827, can be understood as the relatively heavy grain experiencing the sum of many uncorrelated kicks of the chaotic much lighter water molecules. It has become evident that Brownian motion and its associated normal diffusion is too simplistic to describe the variety of diffusion processes in complex systems. There are many situations where the Central Limit Theorem fails \cite{KlafterEtAl96,GaspardWang88,Gouezel04,ContTankov,MantegnaStanley}, and their fluctuations are of the so called L\'evy type rather than of the Gaussian type. Whereas Gaussian processes are continuous processes with finite variance, L\'evy processes (or $\alpha$-stable processes) exhibit jumps of all sizes and have infinite variance. The recent survey articles \cite{MetzlerKlafter00,Sokolov12,HoeflingFranosch13,MetzlerEtAl14,MetzlerEtAl16} discuss a plethora of experimental situations in which anomalous diffusion is observed and provide an overview of current analytical approaches.\\

\noindent
Distilling the information relevant to anomalous diffusion, in particular determining accurately the diffusion rate, $n^s$, presents a significant scientific challenge.  Here, $s=\frac12$ denotes normal diffusion and $s\neq\frac12$ denotes anomalous diffusion. Of particular interest is the case of $\alpha$-stable processes with superdiffusive rate $s=\frac{1}{\alpha}>\frac12$.  In the case of {\em{i.i.d.}}\ data, this problem is well-understood and various techniques such as maximum likelihood methods \cite{DuMouchel73,Nolan01}, quantile matching \cite{McCulloch86} and linear regression of the empirical characteristic function \cite{Koutrouvelis80,Koutrouvelis81} are very effective for determining $\alpha$ and hence $s$; see for example the exposition in \cite{Nolan01}. However, these methods are not designed to deal with data that is noisy and/or non-{\em{i.i.d.}} In practice, the nature of a given time series (which may be {\em{i.i.d.}}, noisy, or even deterministic \cite{GaspardWang88,Gouezel04}) is not known in advance. Hence it is of great importance to have a method that applies to time series regardless of their origin.\\

\noindent
One such method involves the analysis of the mean-square displacement which grows linearly for normal diffusion and sub-linearly or super-linearly for anomalous sub- and super-diffusion, respectively. We show numerically that estimating the asymptotic growth rate of the mean-square displacement is not an efficient method for distinguishing anomalous from normal diffusion; in finite-size time series the statistical behaviour of rare large jumps is not resolved. We therefore suggest to use lower-order moments of order $q\ll1$ where the many well-resolved small jumps contribute more than the rare large jumps. \\

\noindent
In addition it is well known that the estimation of asymptotic growth rates often suffers from a bias caused by a non-zero mean of the observables, and requires error prone pre-processing of the data to subtract the mean, or the employment of detrended fluctuation analysis \cite{PengEtAl94,PengEtAl95}. We propose a new method where an eventual non-zero mean is inherently removed by calculating the $q$'th moments not of the time series directly but of a related twisted time series obtained by rotating the original data with a deterministic periodic driver.\\ A different approach to detect anomalous diffusion is to employ the $p$-variation which recently found lots of application in successfully detecting anomalous behaviour in time series \cite{MagdziarzEtAl09,BurneckiWeron10,MagdziarzKlafter10,HeinImkellerPavlyukevich,BurneckiEtAl12}. 
In this method, however, contamination of the data with additive noise has been shown to mask underlying anomalous diffusive behaviour as discussed in \cite{JeonEtAl13}. 
We compare our method with the standard method of estimating the asymptotic growth rate of the $q$'th moment as well as with the method of $p$-variation and various other standard methods, such as the maximum likelihood method \cite{DuMouchel73,Nolan01}, quantile matching \cite{McCulloch86} and linear regression of the empirical characteristic function \cite{Koutrouvelis80,Koutrouvelis81}, for uncontaminated data and for data contaminated by  biased additive measurement noise. We use data generated from {\em{i.i.d.}}\ random variables as well as from deterministic weakly chaotic maps.\\ 


\noindent
The paper is structured as follows. In Section~\ref{sec.ts} we construct time series exhibiting anomalous diffusion. In Section~\ref{sec.scaling} we present numerical results of the asymptotic growth rates of the $q$'th moment and show that high moments such as the mean-square displacement are not well suited to detect anomalous diffusion in a quantitative way. Section~\ref{sec.methods} briefly describes the standard methods for {\em{i.i.d.}}\ data, the methods to estimate the asymptotic growth rate of the low moments, including our new method, and two versions of the $p$-variation method. Numerical results are presented in Section~\ref{sec.num}. We conclude in Section~\ref{sec.sum} with a summary and discussion.

\section{Time series data}
\label{sec.ts}
\begin{figure*}[!htbp]
\centering
\includegraphics[width = 0.475\columnwidth, height = 6cm]{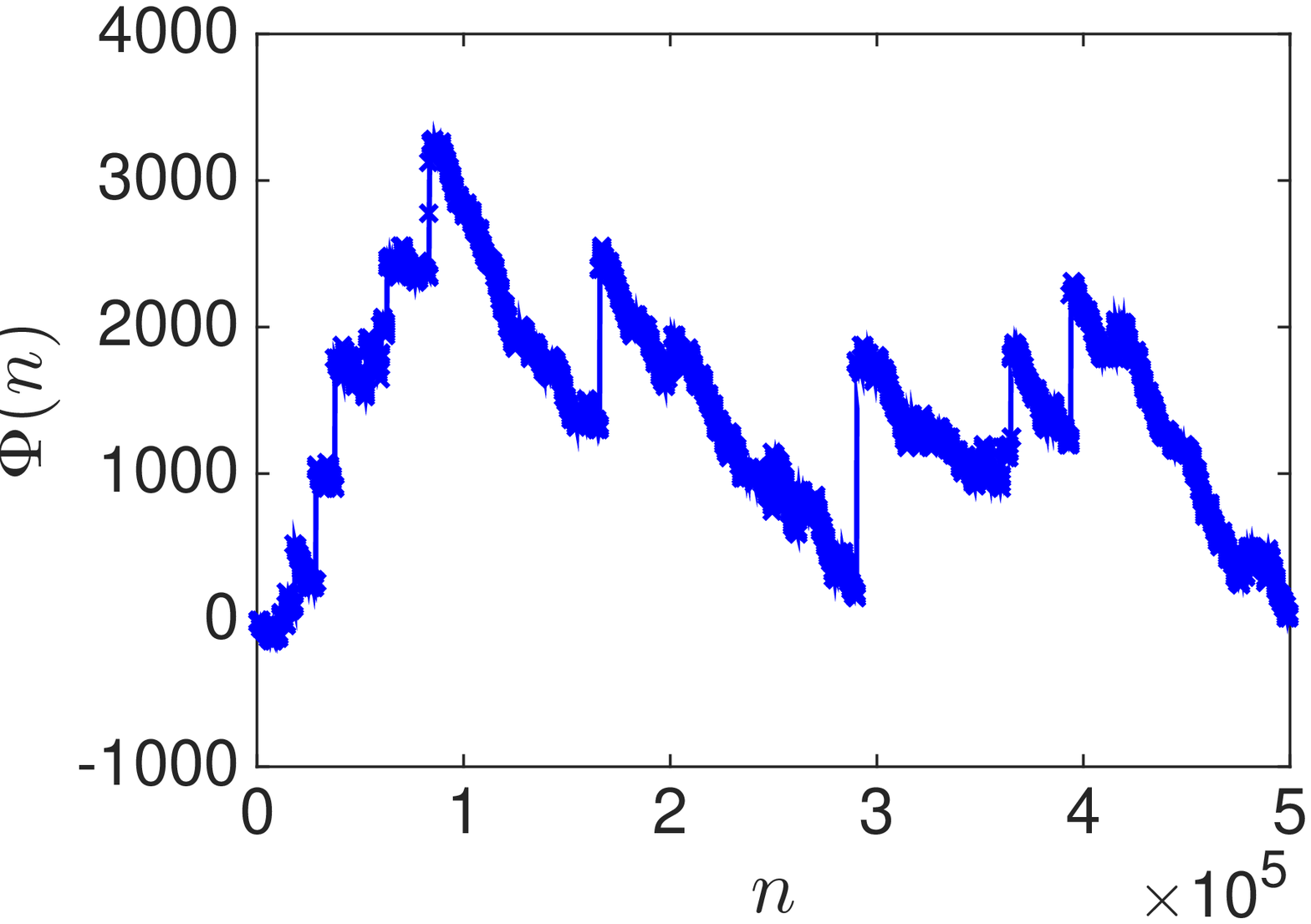}
\includegraphics[width = 0.475\columnwidth, height = 6cm]{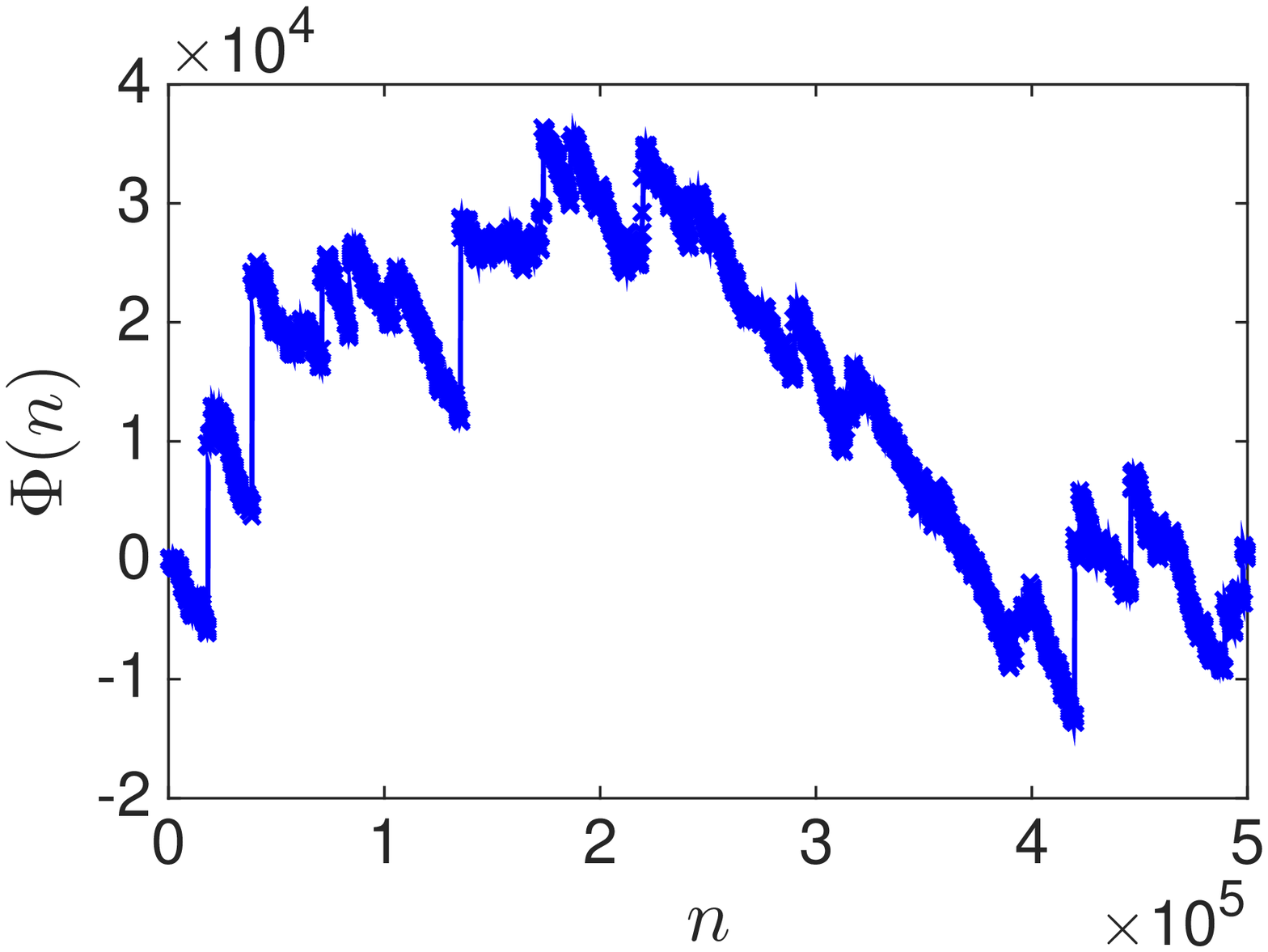}
\caption{Realization of an $\alpha$-stable process $\Phi(n)$ generated from {\em{i.i.d.}}\ variables with $\alpha=1.25$, $\beta=1$, $\mu=1$, $\sigma=0.1$ (left) and through the deterministic map (\ref{e.PM}) with $z=1/1.25=0.8$ and observable $\varphi(y) = 1+y$. The sample mean has been subtracted from the observables $\varphi(j)$.}
\label{fig.ts}
\end{figure*}

We will apply the tests to discrete time series $\{ \varphi(j) \}_{j=1,\cdots,N}$ which are generated both stochastically and deterministically. To distill information about the diffusive nature of the underlying system we construct from the time series the Birkhoff sums
\begin{align}
\Phi(n) = \sum_{j=0}^{n-1}\varphi(j)\;.
\label{e.BS}
\end{align}
In the stochastic case, we consider {\em{i.i.d.}}\ sequences of $\alpha$-stable random variables $\varphi(j)$. Such random variables $S_\alpha(\beta,\mu,\sigma)$ are uniquely characterized by four parameters: asymmetry parameter $\beta$, location parameter $\mu$ and spread parameter $\sigma$ together with $\alpha$.
Numerically, we generated these random variables via the method of Chambers, Mallows and Stuck \cite{ChambersMallowsStuck76}.
In Figure~\ref{fig.ts} we show $\Phi(n)$ for $\alpha=1.5$, $\beta=1$, $\mu=1$ and $\sigma=0.1$.
The linear drift in the Birkhoff sum $\Phi(n)$ caused by $\mu \neq 0$ has been subtracted by computing the sample mean for $\alpha>1$, i.e.\ by considering $\varphi(j) \to \varphi(j) - (1/n)\sum^{n-1}_{m=0} \varphi(m)$.\\ To generate the time series deterministically we employ Pomeau-Manneville intermittency  maps~\cite{PomeauManneville80}. In particular, we use the map $y_{n+1}=f(y_n)$ with $f:[0,1]\to[0,1]$ studied by~\cite{LiveraniEtAl99}
\begin{align}
f(y)=
\begin{cases} 
y(1+2^z y^z), & y\in[0,\frac12)\\ 
2y-1, & y\in[\frac12,1] 
\end{cases}
. 
\label{e.PM}
\end{align}
This map has a neutral fixed point at $y=0$. For $z=0$ the map reduces to the doubling map which preserves the uniform measure on the interval $[0,1]$ and exhibits exponential decay of correlations. For $z\in(0,1)$, there exists a unique absolutely continuous invariant ergodic probability measure, and correlations decay at the rate $n^{-(z^{-1}-1)}$ \cite{Hu04}. Correlations are summable if and only if $z<\frac12$, and in this situation the central limit theorem applies with $n^{-\frac12}\Phi(n)$ converging in law to a normal distribution for mean zero H\"older observables $\varphi$ \cite{Young99}. 
For $z\in(\frac12,1)$, however, Gou\"ezel~\cite{Gouezel04} proved that for sufficiently smooth mean zero observables $\varphi(y)$ which are non-zero at the neutral fixed point, the central limit theorem fails and instead $n^{-z}\sum_{j=0}^{n-1}\varphi(y_j)$ converges in distribution to a stable law of exponent $\alpha=1/z$, asymmetry $\beta=\pm1$ and mean
$\mu=0$. The jumps are produced by the orbit spending prolonged times near $0$ with $\varphi\approx \varphi(0)$. In order to get better statistics, we consider the induced map, which effectively condenses the many small jumps to a single big jump. The inducing is performed by passing from the nonuniformly expanding map $f:[0,1]\to[0,1]$ to the uniformly expanding first return map $F=f^\tau:Y\to Y$ with $Y=[1/2,1]$ where $\tau(y)=\inf\{n\ge1:f^ny\in Y\}$ is the first return time back into the set $Y$ for $y\in Y$. Induced observables are then defined as
\begin{align}
\varphi_I(y)=\sum_{\ell=0}^{\tau(y)-1}\varphi(f^\ell y) ,
\end{align}
leading to $\Phi(n)=\sum_{j=0}^{n-1}\varphi_I\circ F^j$ via iteration of this
procedure. 
 
In Figure~\ref{fig.ts} we show the time series $\Phi(n)$ for $\alpha=1.25$ generated via the map (\ref{e.PM}) with $z=0.8$ for the observable $\varphi(y)=1+y$. The linear drift of the Birkhoff sum $\Phi(n)$ was again approximately eliminated by subtracting the sample mean.\\

%


\section{Scaling behaviour of the $q$'th moments}
\label{sec.scaling}
We now investigate the scaling behaviour of the $q$'th moment $\E|\Phi(n)|^q$. Envoking ergodicity the $q$'th moment is expressed by the time average 
\begin{align}
\E|\Phi(n)|^q = \lim_{N\to\infty}\frac1N\sum_{j=0}^{N-1}|\Phi(j+n)-\Phi(j)|^q\;.
\label{e.Eq}
\end{align}
In the case of zero-mean {\em{i.i.d.}}\ $\alpha$-stable random variables, the $q$'th moments exist for $q<\alpha$ and scale as
\begin{align}
\E|\Phi(n)|^q \sim
n^\frac{q}{\alpha} .
\label{e.ESn_iid}
\end{align}
For $q\ge\alpha$ the $q$'th moments do not exist. In the case of anomalous diffusion of underlying deterministic weakly chaotic dynamics, the moments exist for all values of $q$ and scale as follows 
\begin{align}
\begin{cases}  \E|\Phi(n)|^q \sim n^{q/\alpha} & q<\alpha \\
\E|\Phi(n)|^q \approx n^{q+1-\alpha} & q>\alpha \end{cases}.
\label{e.ESn}
\end{align}
(We write $a_n\sim b_n$ if there exists a constant $c>0$ such that $\lim_{n\to\infty} a_n/b_n=c$.  We write $a_n\approx b_n$ if there exists constants $C_1,C_2>0$ such that $C_1\le a_n/b_n\le C_2$ for all $n\ge1$.) For the Brownian motion case we obtain the linear scaling of the mean-square displacement $\E|\Phi(n)|^2\sim n$. Bi-linear scaling as in (\ref{e.ESn}) was experimentally observed in active transport of polystyrene beads in living cells \cite{GalWeihs10} and has been studied theoretically in infinite horizon billiards, intermittent maps and L\'evy walks \cite{ArmsteadEtAl03,CourbageEtAl08,ArtusoCristadoro03,RebenshtokEtAl14}. For a rigorous mathematical proof of (\ref{e.ESn}), see  \cite{MelbourneGouezel14,DedeckerMerlevede15}.\\

We now investigate the scaling behaviour of the $q$'th moments by plotting the growth rate
\begin{align}
\gamma(q) = \lim_{n\to \infty} \frac{\log(\E|\Phi(n)|^q)}{\log(n)}
\label{e.qthM}
\end{align}
for several values of $q$ for {\em{i.i.d.}}\ observables and for observables obtained from a deterministic intermittency map. To avoid any issue with a non-zero mean of the observables creating non-negligible drift terms, we symmetrize the intermittency map (\ref{e.PM}) and consider the map $y_{n+1}=f_{\rm{sym}}(y_n)$ with $f_{\rm{sym}}:[-1,1]\to[-1,1]$
\begin{align}
f_{\rm{sym}}(y) = 
\begin{cases} 
-2y, & 0\le y\le\frac12 \\
1-(1-y)(1+2^z(1-y)^z), & \frac12\le y\le1 \\
-f_{\rm{sym}}(-y), & -1\le y\le 1
\end{cases}
\label{e.PMsym}
\end{align}
with neutral fixed points at $y=\pm1$. 
To determine the asymptotic growth rate $\gamma(q)$ from a single time series, we need to respect the double limit in the temporal average (\ref{e.Eq}). The double limit requires us to choose $n\ll N$. In practice we use $n\le N/10$. The asymptotic growth rate is then determined by linear regression of $\E|\Phi(n)|^q$.

\begin{figure*}[!htbp]
\centering
\includegraphics[width = 0.475\columnwidth, height = 6cm]{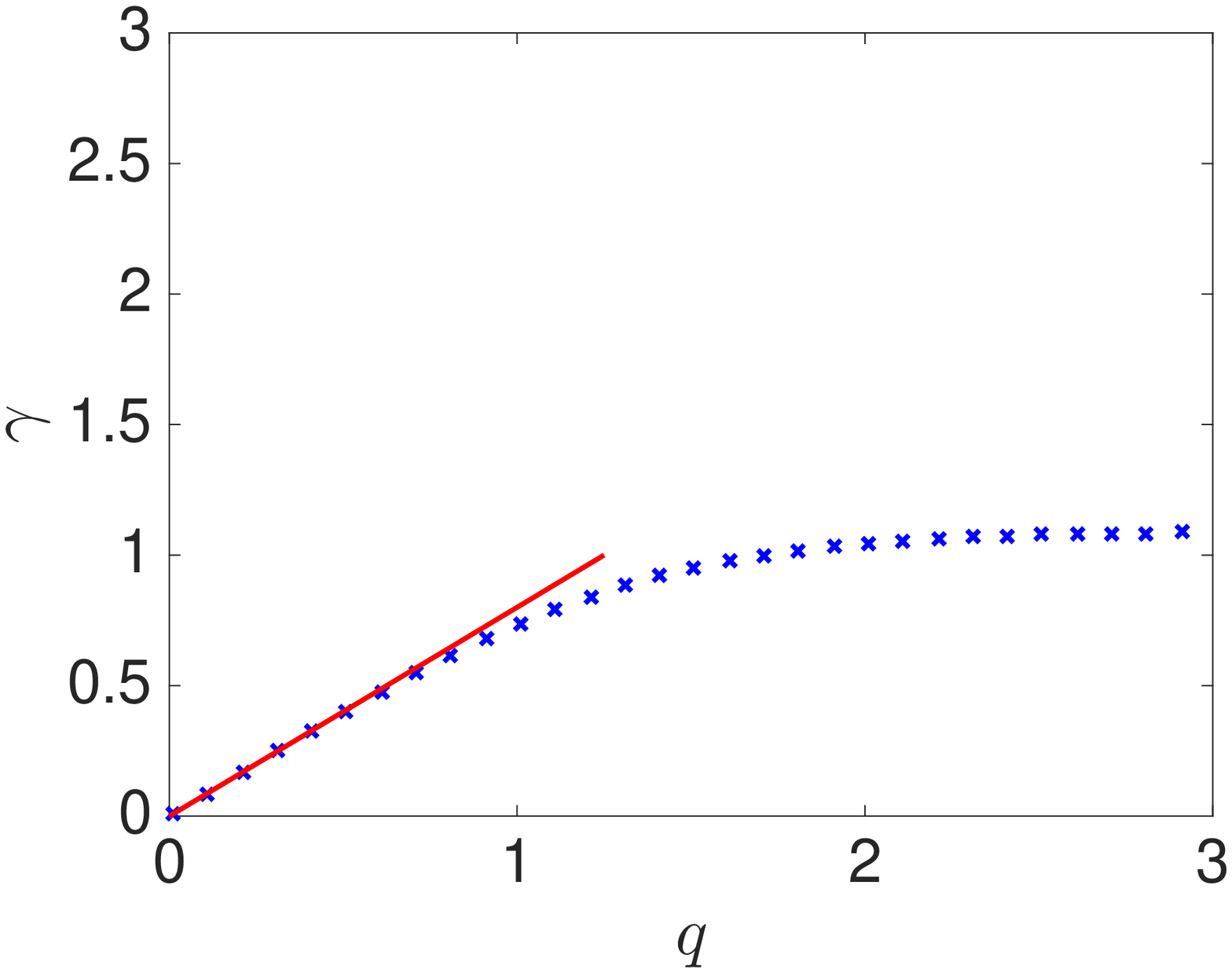}
\includegraphics[width = 0.475\columnwidth, height = 6cm]{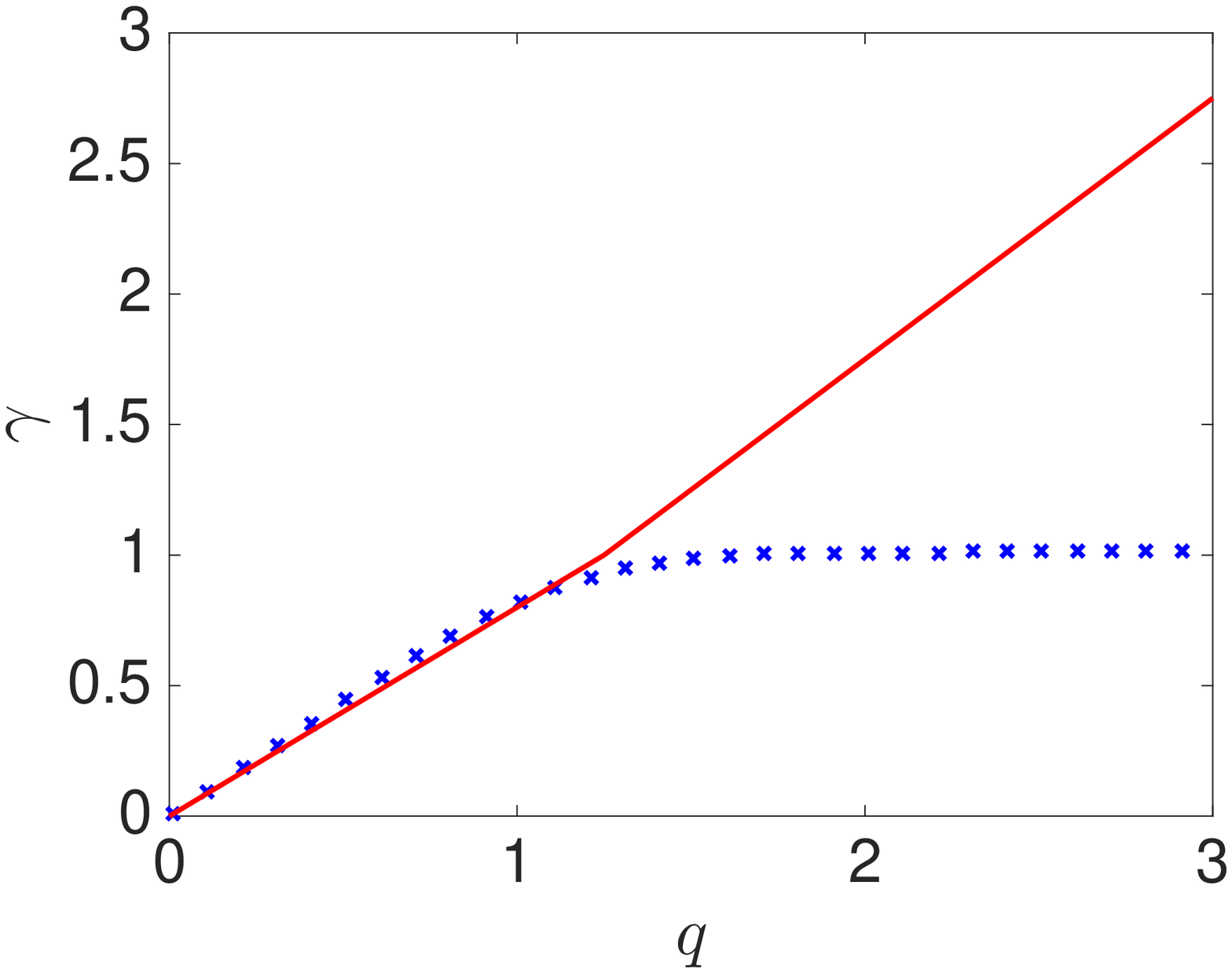}
\caption{The growth rate $\gamma$ of the $q$'th moment $\mathbb{E}|\Phi|^q \sim n^\gamma$ for $\varphi(y) = y$ as a function of $q$ for the {\em{i.i.d.}}\ case with $\mu=0$, $\sigma=0.1$, $\beta=0$ and $\alpha=1.25$ (left) and for the Pomeau-Manneville map (\ref{e.PMsym}) (right) for $\alpha=1.25$. The continuous curve (online red) depicts the theoretical result according to (\ref{e.ESn_iid}) or (\ref{e.ESn}), respectively.} 
\label{fig.qthE}
\end{figure*}

\begin{figure*}[!htbp]
\centering
\includegraphics[width = 0.5\columnwidth, height = 6cm]{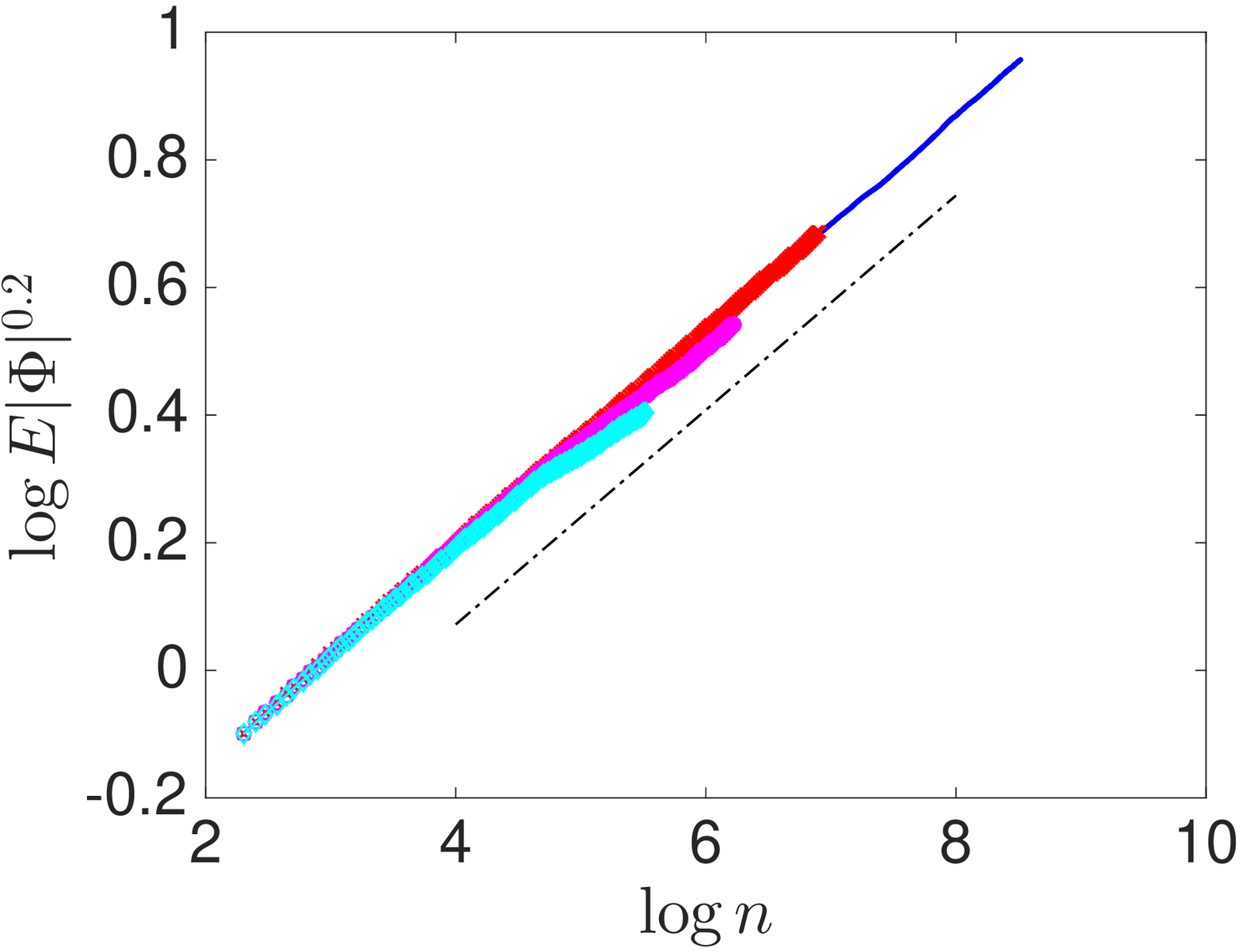}
\includegraphics[width = 0.475\columnwidth, height = 6cm]{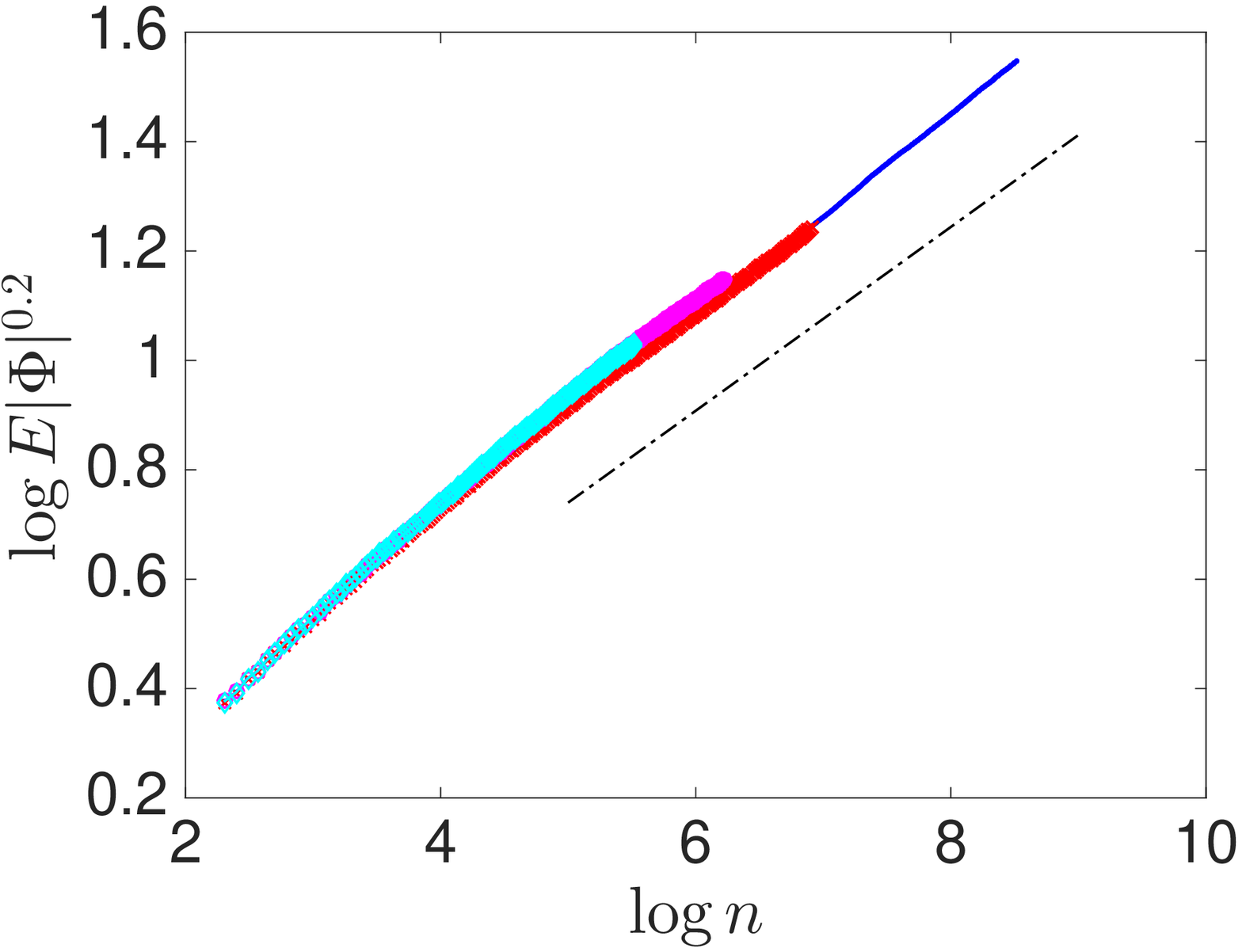}\\
\includegraphics[width = 0.475\columnwidth, height = 6cm]{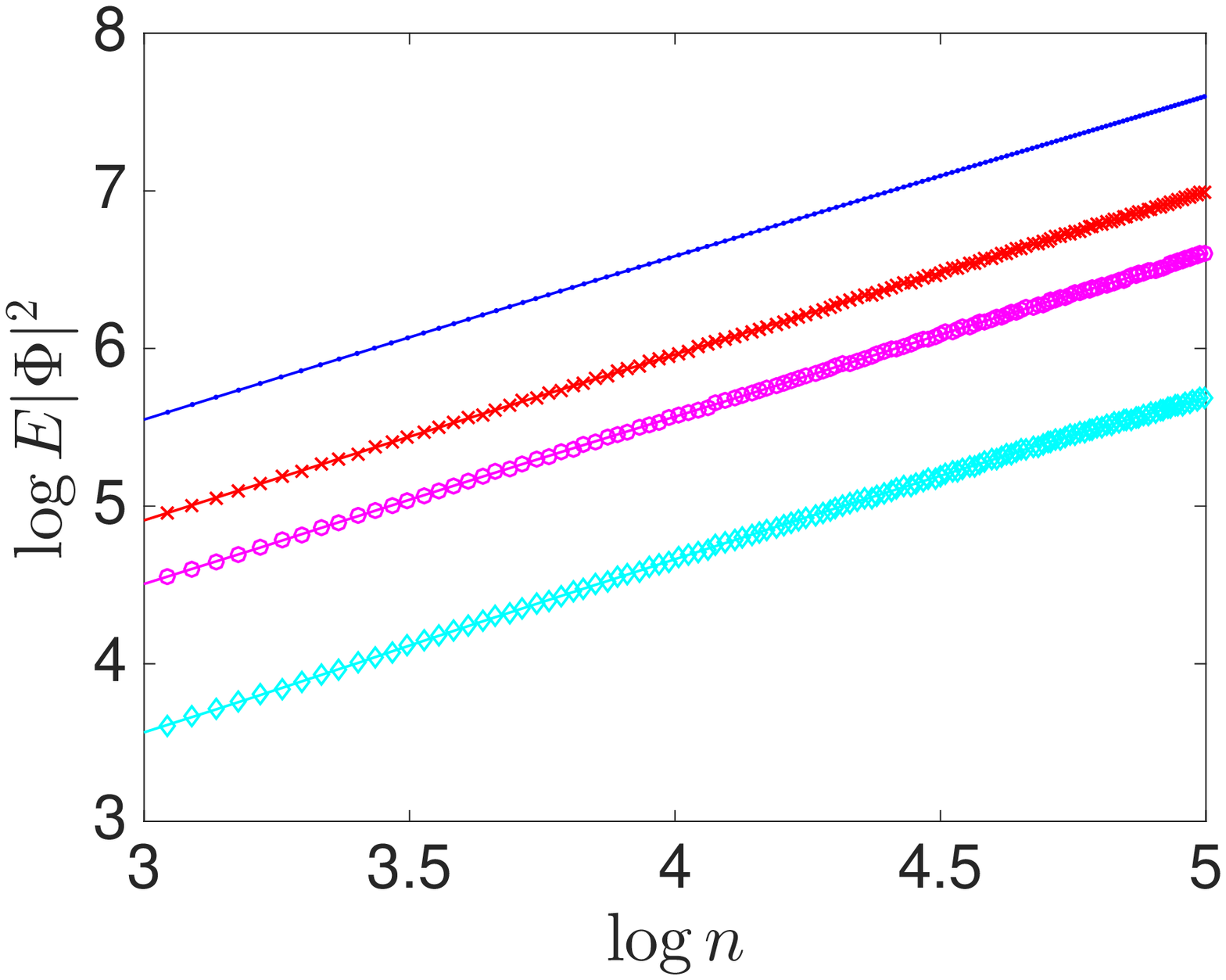}
\includegraphics[width = 0.475\columnwidth, height = 6cm]{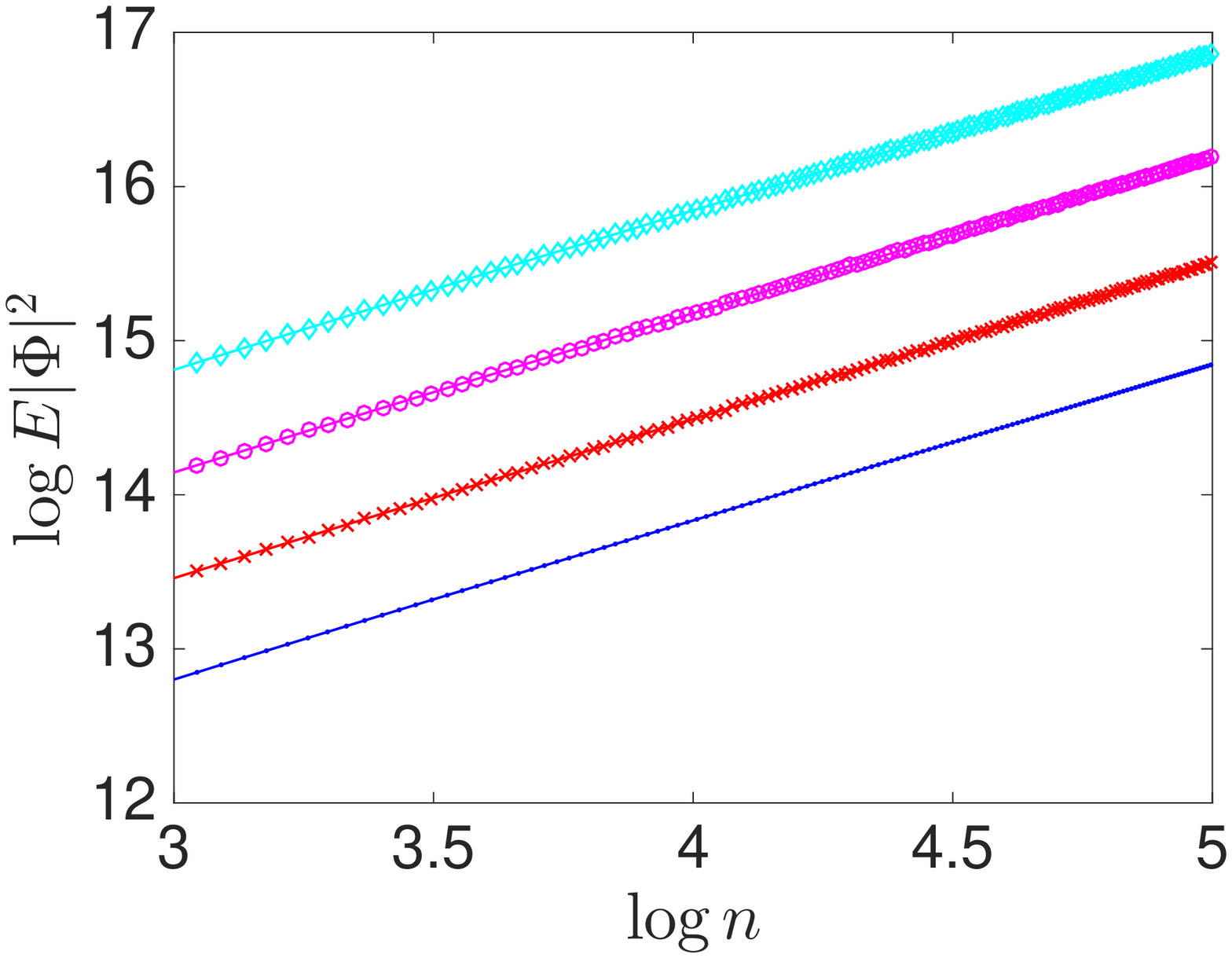}
\caption{The $q$th moment $\mathbb{E}|\Phi|^q$ for $\varphi(y) =y$ for the {\em{i.i.d.}}\ case (left) and for the Pomeau-Manneville map (\ref{e.PMsym}) (right). Results are shown for $\alpha=1.25$ for time series of length $N=500,000$ (dashed line, online blue), $N=100,000$ (crosses, online red), $N=50,000$ (open circles, online magenta) and $N=25,000$ (diamonds, online cyan). Top: For $q=0.2<\alpha$. Bottom: For $q=2>\alpha$. The dashed lines in the top figures show the theoretical slope as calculated via (\ref{e.ESn_iid}) and (\ref{e.ESn}), respectively. The slope in the bottom figures is approximately $1$ (cf. Figure~\ref{fig.qthE}).}
\label{fig.qthE2}
\end{figure*}

Figure~\ref{fig.qthE} shows results of numerical simulations for time series of length $N=500,000$. Whereas the simulations confirm the theoretical growth rate $\gamma(q)$ implied by (\ref{e.ESn}) for small values of $q$, it is clearly violated for large $q>\alpha$. 
In particular, for the usual value $q=2$, the implied value for the anomalous diffusion is $\alpha_{\rm est}=2/\gamma\approx 2$. This suggests that the estimation of the mean-square displacement ($q=2$) would falsely classify anomalous diffusion as normal with a linear growth.\\ In Figure~\ref{fig.qthE2} we show the $q$'th moment as a function of $n$ for several values of $N$ for $q=0.2$ and $q=2$ for $\alpha=1.25$. For $q=0.2$, the convergence to the theoretical scaling result (\ref{e.ESn_iid}) and (\ref{e.ESn}), respectively, is clearly seen (top panel). For $q=2$ (bottom panel), the growth rate is approximately equal to $1$ for the {\em{i.i.d.}}\ case as well as for the Pomeau-Manneville case, consistent with the results presented in Figure~\ref{fig.qthE}. It is also clearly seen that the $2$nd moments have not converged. Note that this is consistent with the nonexistence of the $2$nd moment in the {\em{i.i.d.}}\ case. For the deterministic Pomeau-Manneville case in which the $2$nd moment exists, however, this illustrates that $N=500,000$ is insufficient to determine the slope of $3-\alpha$ (cf (\ref{e.ESn})).\\

\noindent
The results show that calculating the mean-square displacement is not satisfactory for distinguishing anomalous superdiffusion and normal diffusion in finite time series; note that the time series of $N=500,000$ data points is rather large. The results rather suggest to use lower moments with small values of $q$ to estimate the anomalous coefficient $\alpha$. 
A heuristic explanation for the superior performance of lower moments is that in a finite data set the statistics of the large jumps are necessarily not well resolved. For low values of $q\ll 1$ the smaller jumps, for which better statistics are available within a finite data set, receive a relatively larger weighting than larger jumps in the time average~(\ref{e.Eq}). The relative importance of large jumps in the $q$'th moment (\ref{e.Eq}) is increased for large values of $q$.

%


\section{Methods of detection}
\label{sec.methods}

\subsection{Standard estimation methods}
\label{sec.nolan}
Parameter estimation for $\alpha$-stable distributions is well-developed for the case of {\em{i.i.d.}}\ random variables which are not contaminated by noise. There are numerous techniques such as maximum likelihood estimators \cite{DuMouchel73,Nolan01}, quantile matching \cite{McCulloch86} and linear regression of the empirical characteristic function \cite{Koutrouvelis80,Koutrouvelis81}. The reader is referred to \cite{Weron95,Nolan01} for a detailed description and for numerical comparisons in the case of pure {\em{i.i.d.}}\ random variables. In the numerical results presented in Section~\ref{sec.num} we use publicly available matlab routines for the quantile matching \cite{BorakWeronMatlab,VeilletteMatlab} and for the linear regression method \cite{VeilletteMatlab}, and use the software package {\rm{STABLE}} \cite{NolanStable} for the maximum-likelihood estimator\footnote{We have also used the matlab built-in command {\em{mle}} \cite{MATLAB} for the maximum-likelihood estimator, but found it less reliable than the command {\em{stablefitmle}} from the {\rm{STABLE}} package.}.

\subsection{Measuring the asymptotic growth rate of the $q$'th moment}
\label{sec.qm}
The first method is the standard determination of the asymptotic growth rate $\gamma$ of the $q$'th moment via linear regression for a given time series of length $N$.  Motivated by the numerical results from the previous Section we choose $q=1/8$. 
A non-zero mean of the observables $\varphi(j)$ would dominate the asymptotic behaviour of the $q$'th moments leading to $\gamma(q)=q$, independent of the underlying diffusive nature of the dynamics. We therefore subtract the sample mean $N^{-1}\sum_{j=0}^{N-1}\varphi(j)$ from the observables for $\alpha>1$. Note that the mean is not defined for $\alpha<1$ in the case of {\em{i.i.d.}}\ random variables (cf. Section~\ref{sec.scaling}). Hence, without {\em{a priori}} knowledge of $\alpha$, subtracting the sample mean is problematic.


\subsection{Measuring the asymptotic growth rate of the twisted $q$'th moment}
To account for a possible non-zero mean of the observables $\varphi(j)$ we consider instead of (\ref{e.BS}) the following rotated Birkhoff sum
\begin{align}
\Phi_c(n) = \sum_{j=0}^{n-1}\varphi(j)\cos c j\;
\label{e.BS2}
\end{align} 
where $c\neq 0$ is fixed. Including the rotational variable $\cos cj$ assures that the mean of $\Phi_c$ is automatically zero. (A rigorous justification is based on~\cite[Section 3]{NicolMelbourneAshwin01} via the ergodic theorem.  Intuitively, the linear drift of the Birkhoff sum $\Phi_c(n)$ has no preferred direction in the complex plane due to the rotation variable, and hence averages to zero. 
The inclusion of a rotational variable has proven very useful in the detection of deterministic chaos using the $0$-$1$ test for chaos \cite{GottwaldMelbourne04,GottwaldMelbourne05,GottwaldMelbourne09,GottwaldMelbourne09b}.)
We will see in Section~\ref{sec.num} that this has advantages over manually subtracting the sample mean, as in the Section~\ref{sec.qm}, which may contaminate the statistics. We then calculate the $q$'th moment $\mathbb{E}|\Phi_c|^q$ and measure its asymptotic growth rate $\gamma_c$. We again use $q=1/8$. Possible resonances between the rotational variable and the actual underlying dynamics encoded in the observable will obscure the information contained in $\gamma_c(q)$. Such resonances correspond to a term in the Fourier decomposition of the time series $\varphi(j)$ proportional to $\exp(-i c k)$ leading to $\Phi_c(n)\sim n$ implying an asymptotic growth rate $\gamma_c = q$, independent of the actual diffusive behaviour. We therefore compute the median of $\gamma_c$ over several values of $c$. 
We choose to compute the median rather than the mean to avoid the effect of outliers. In practice we find that $100$ randomly chosen values of $c\in(\pi/5 ,4 \pi/5)$ are sufficient. 

\subsection{$p$-variation method}
The $p$-variation associated with a process $\Phi$ is defined as the asymptotic limit 
\begin{align}
V_p(t) = \lim_{n\to\infty}V_p^n(t)\; ,
\end{align}
where $V_p^n(t)$ is the partial sum of increments of the observable $\Phi(n)$ given by
\begin{align*}
V_p^n(t) = \sum_{k=1}^{\lfloor nt\rfloor }\Big|\Phi\Big(\frac{k}{n}\Big) - \Phi\Big(\frac{k-1}{n}\Big)\Big|^p.
\end{align*} 
For $p=1$, $V_1(t)$ reduces to the total variation, and for $p=2$, $V_2(t)$ reduces to the quadratic variation. It is known that for Brownian motion, $V_2(t)\sim t$ and $V_p(t)=0$ for any $p>2$. In the case of subdiffusion, the $p$-variation allows to distinguish fractional Brownian motion and Continuous Time Random Walk (CTRW) diffusion \cite{MagdziarzEtAl09,MagdziarzKlafter10}. For fractional Brownian motion, $V_2(t)=\infty$ and $V_{2/\gamma^\star}(t)\sim t$, whereas for CTRW, $V_2(t)$ is a monotonically increasing step function and $V_{2/\gamma^\star}(t)=0$, where $\gamma^\star=\gamma(2)$ is the asymptotic growth rate of the mean-square displacement. For superdiffusion, $V_p^n(t)$ converges for $p>\alpha$ and diverges for $p<\alpha$ as $n\to \infty$. This suggests to estimate $\alpha$ by determining the smallest value $p^\star$ for which convergence occurs and set $\alpha_{\rm{est}}=p^\star$. In practice, we subsample a time series of length $N$ into $2^m$ data points with equal spacing $N/2^m$ with $m=0,\cdots,\lfloor \log N/\log 2 \rfloor$. For the finest samplings we estimate a linear approximation $\hat V_p^n(t)$ by linear regression of $V_p^n(t)$.  We then determine $p^\star$ as the minimal value of $p$ for which the $\ell_1$-norm of the difference between two consecutive samplings $|\hat V_p^n(t) - \hat V_p^{n-1}(t)|$ falls below some threshold $\theta_p$. The choice of the threshold $\theta_p$ is, of course, arbitrary and depends on the underlying dynamical system which is analyzed.

\subsection{Modified $p$-variation method}
In \cite{CorcueraEtAl07,HeinImkellerPavlyukevich} theorems were proved showing that for an $\alpha$-stable random variable with location parameter $\mu=0$ and $\alpha=p/2$ (and any values of $\beta$ and $\sigma$) its $p$-variation $V_p^n(t)$ converges in distribution to an $\alpha=1/2$-stable random variable $S_{1/2}(1,0,\sigma)$ with some specified spread parameter $\sigma$. In \cite{HeinImkellerPavlyukevich} this was developed into a time series analysis method using a Kolmogorov-Smirnov test and finding the value of $p=2\alpha$ for which the empirical cumulative distribution function is closest to the target cumulative distribution function of $S_{1/2}(1,0,\sigma)$. To estimate the cumulative distribution function of $V_p^n(t)$, an ensemble of $p$-variations is generated by segmenting the time series into $M$ pieces, each being of length $\lfloor N/M\rfloor$. This tacitly assumes that the samples are uncorrelated which is only approximately true for sufficiently long segments in the deterministic case. The minimal Kolmogorov-Smirnov distance is determined by varying the spread parameter $\sigma$ of the target distribution $S_{1/2}(1,0,\sigma)$ for each value of $p$. The value $p^\star$ for which the minimum is attained then determines $\alpha=p^\star/2$. The precise mathematical statement is provided in the appendix.  
For details on the modified $p$-variation method see \cite{CorcueraEtAl07,HeinImkellerPavlyukevich}.



\section{Numerical results}
\label{sec.num}

We use a time series of length $N=25,000$ calculated from {\em{i.i.d.}}\ random variables and from weakly chaotic deterministic variables. We show results for pure data and for noise-contaminated data. To calculate the $q$'th moments  $\mathbb{E}|\Phi|^q$ we employ $q=1/8$. 
For the $p$-variation method we use $\theta_p=0.01$ and cycle through $p$ in increments of $\Delta p=0.005$. We found that larger values of $\theta_p$ perform better for larger values of $\alpha$ but worse for smaller values of $\alpha$. For the modified $p$-variation method, we choose $M=250$ samples of length $100$ each, and cycle through $10,000$ values of the spread parameter $\sigma\in(10^{-2},10^{10})$ (equidistant in $\log$-space).

\subsection{Results for the {\em{i.i.d.}}\ case}
We use a time series of length $N=25,000$ constructed by the Chambers, Mallows and Stuck \cite{ChambersMallowsStuck76} method. We set the asymmetry parameter $\beta=1$ and the spread parameter $\sigma=0.1$, and allow for a non-zero mean parameter $\mu=2$.  

\begin{figure*}[!htbp]
\centering
\psfrag{$RRR$}[][]{$\mathbb{R}$}
\includegraphics[width = 0.475\columnwidth, height = 6cm]{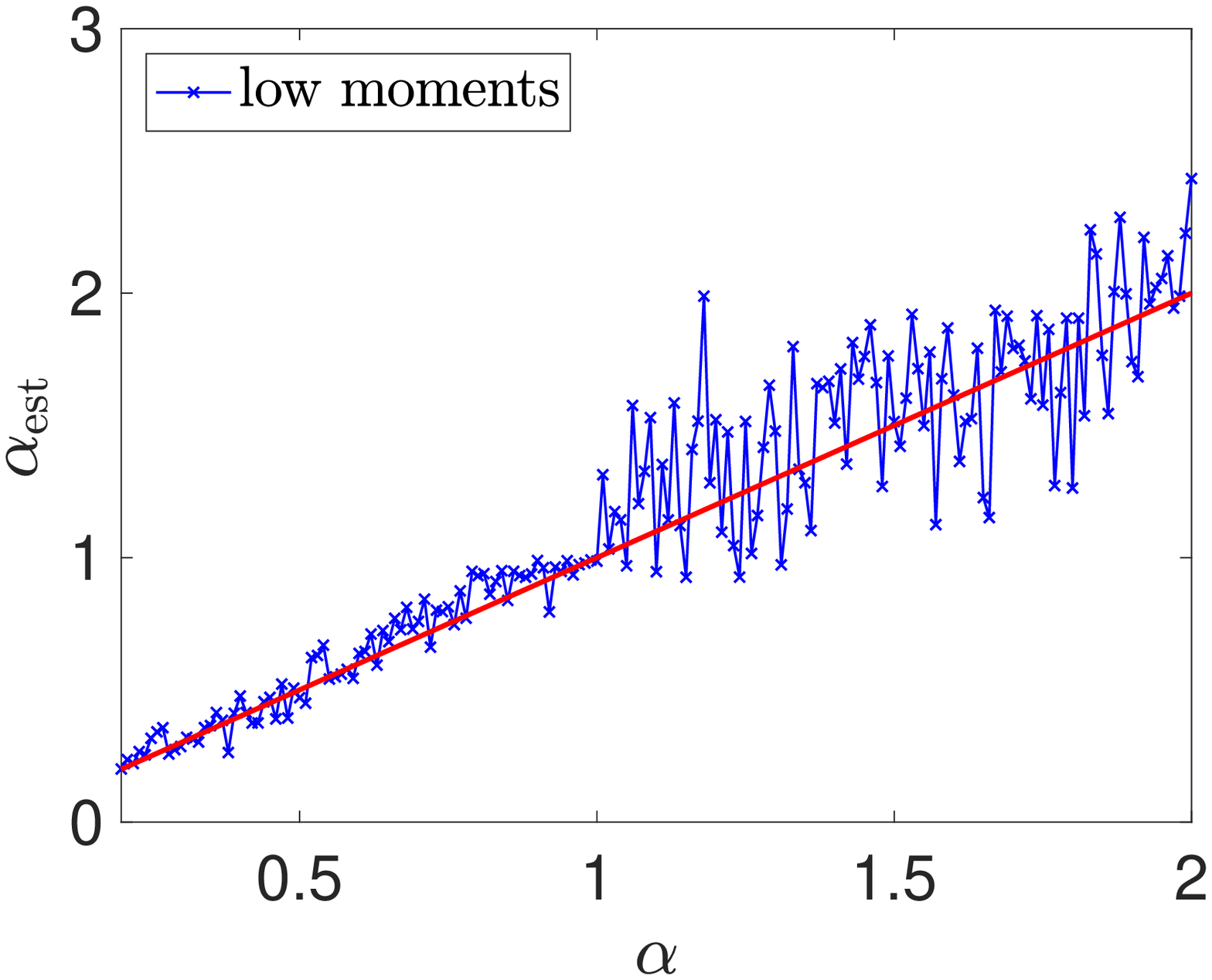}
\includegraphics[width = 0.475\columnwidth, height = 6cm]{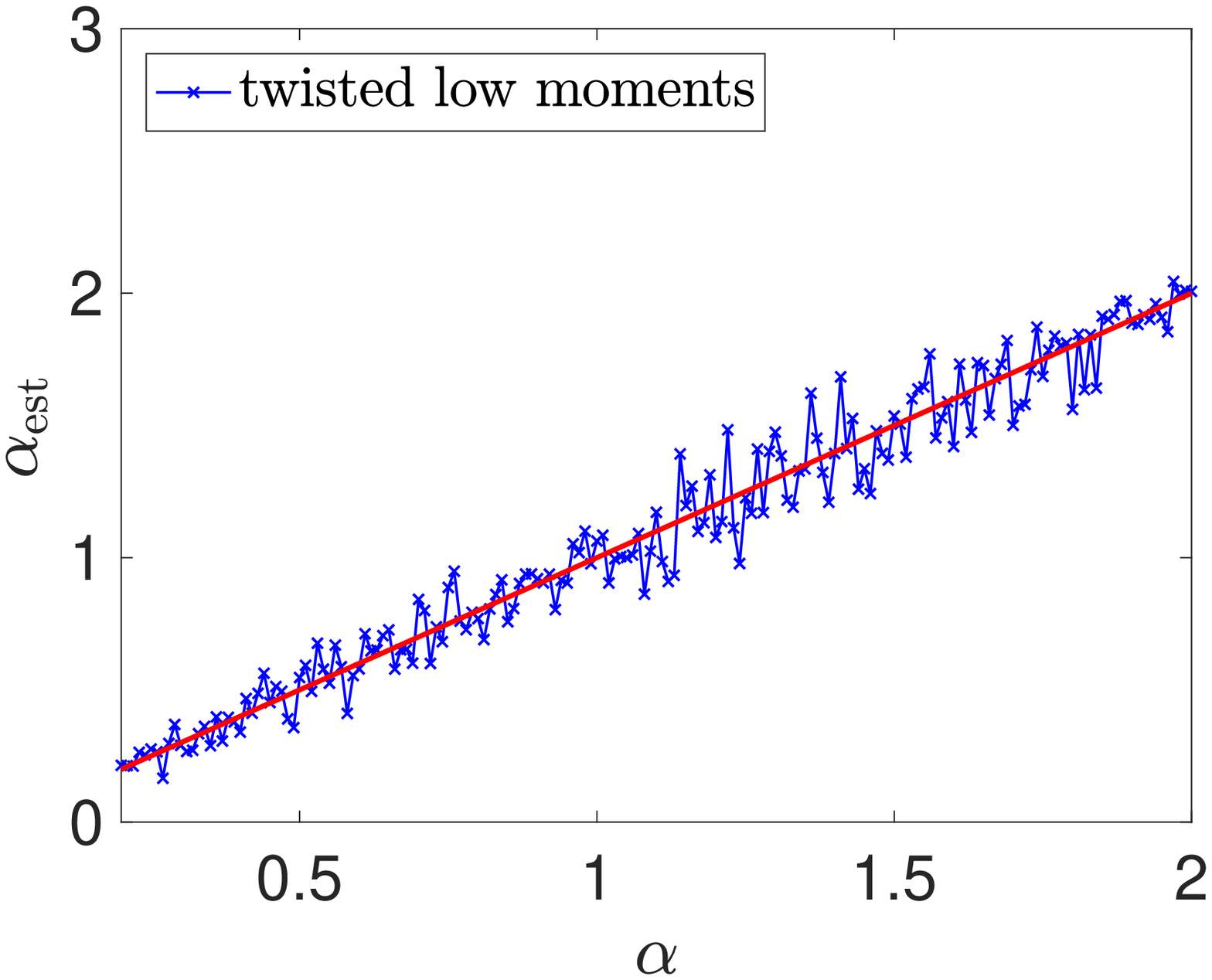}
\\
\includegraphics[width = 0.475\columnwidth, height = 6cm]{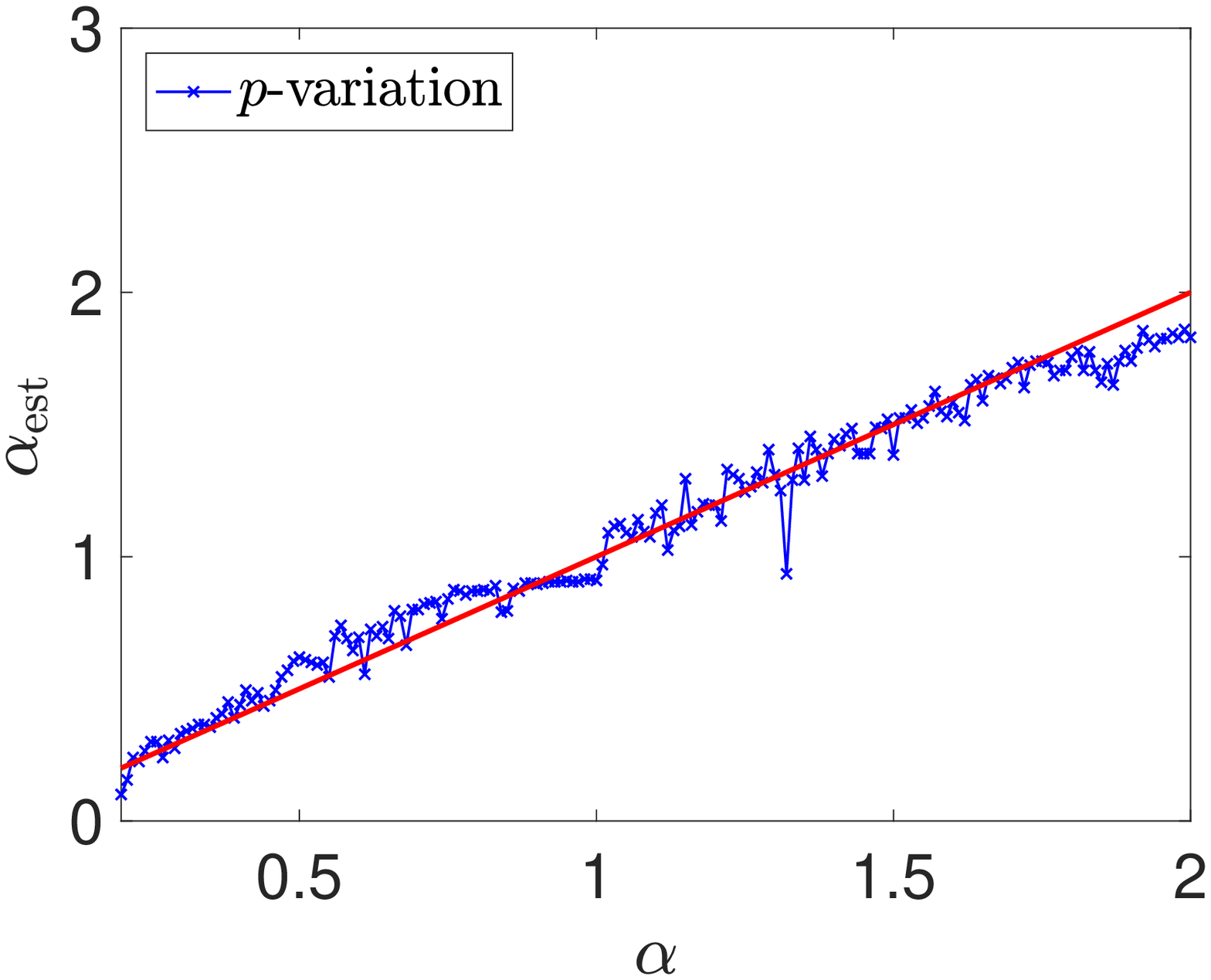}
\includegraphics[width = 0.475\columnwidth, height = 6cm]{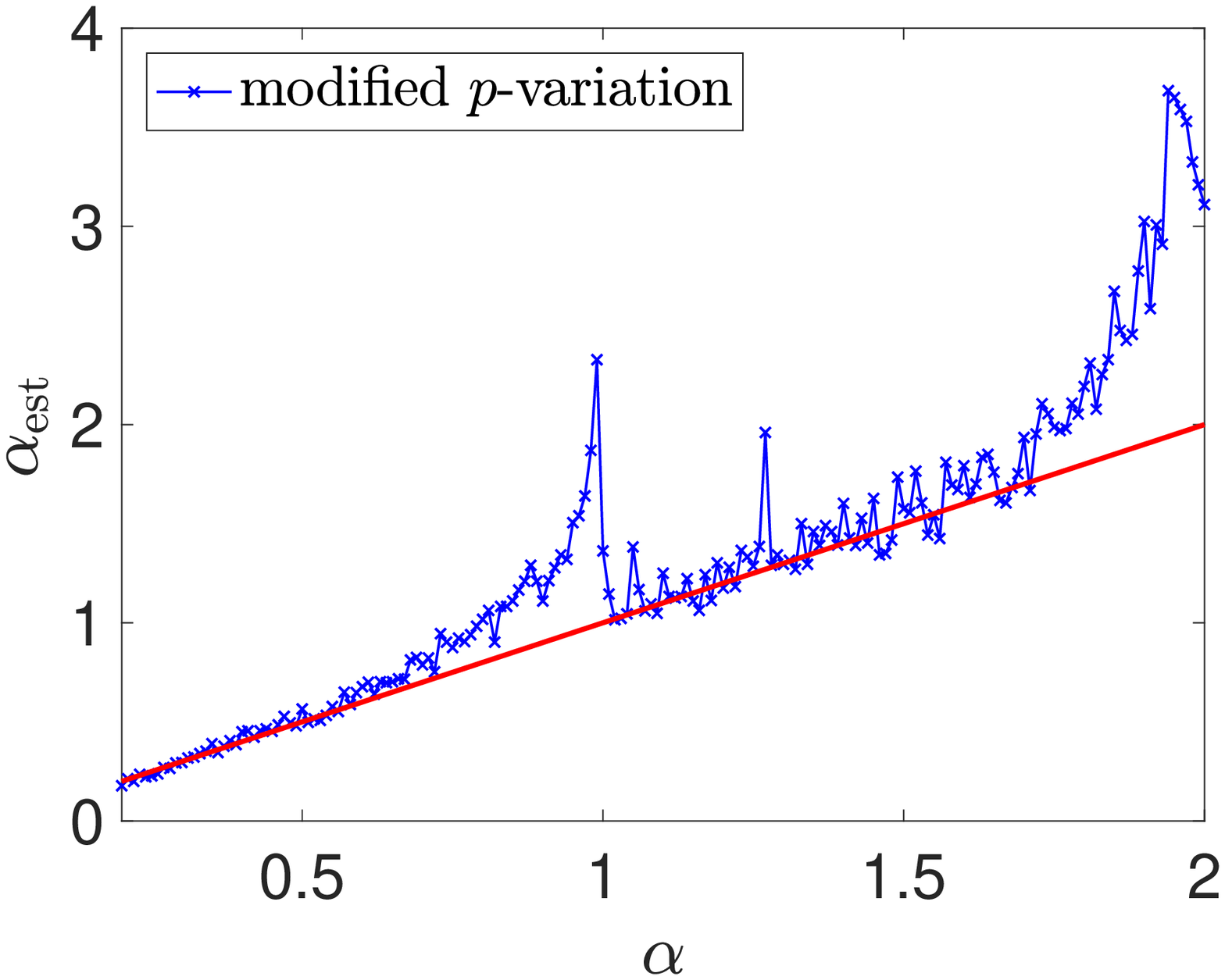}
\\
\includegraphics[width = 0.475\columnwidth, height = 6cm]{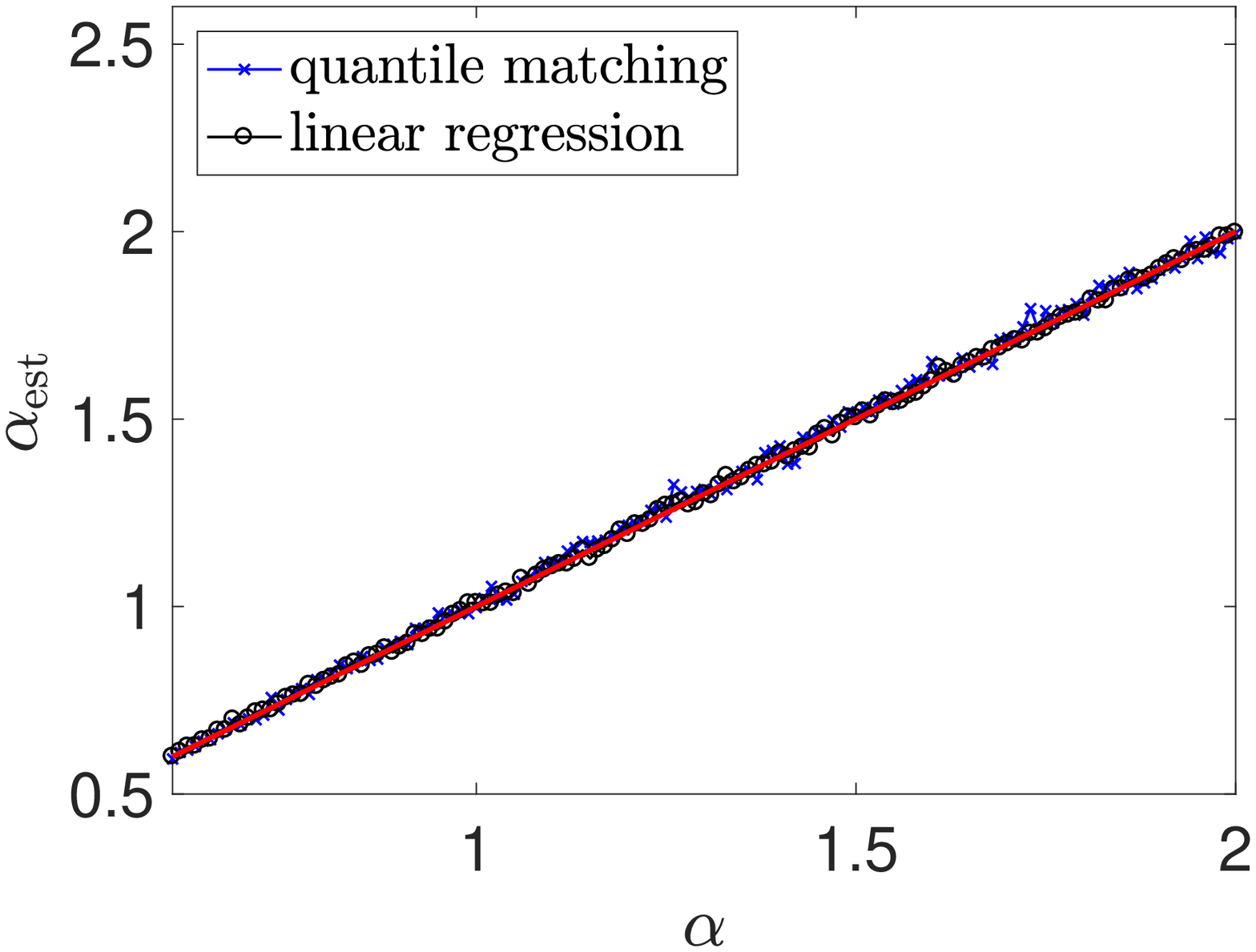}
\includegraphics[width = 0.475\columnwidth, height = 6cm]{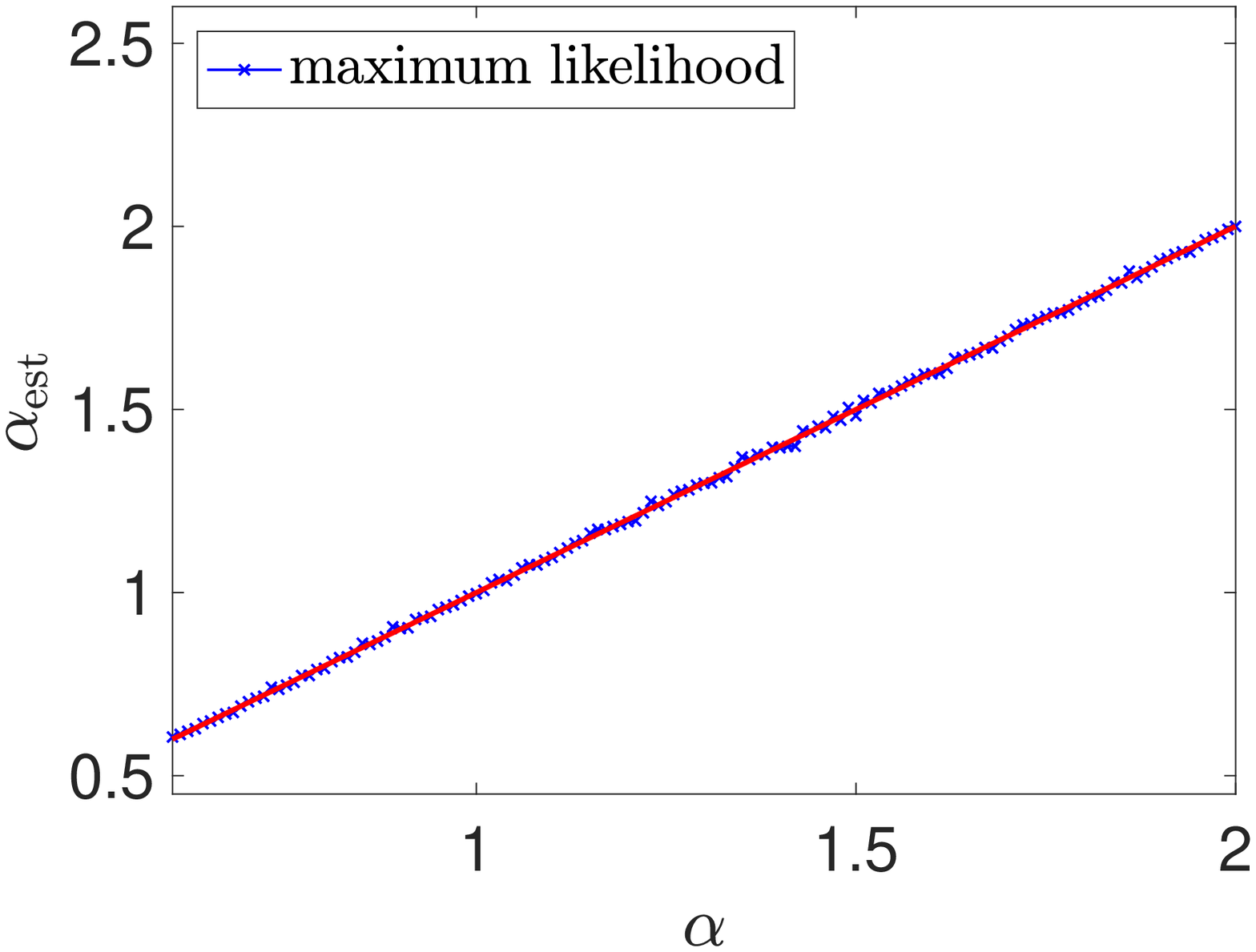}
\caption{Estimates $\alpha_{\rm{est}}$ as calculated for several values $\alpha\in(0.2,2)$ for {\em{i.i.d.}}\ observations. Top left: method of low $q$'th moment with $q=1/8$. Top right: method of twisted low $q$'th moment with $q=1/8$. Middle left: $p$-variation method. Middle right: modified $p$-variation method (note the larger range on the $y$-axis). Bottom left: Methods using the empirical characteristic function. Bottom right: Maximum likelihood estimator. The straight lines show the theoretical answer $\alpha_{est}=\alpha$.}
\label{fig.IID_all}
\end{figure*}

In Figure~\ref{fig.IID_all} we show results for the estimated value of $\alpha$ for the methods 
described in the previous Section. For the methods using the asymptotic growth rate $\gamma$ we estimate the implied value for $\alpha_{\rm{est}} = q/\gamma$ for $\alpha>q$. The method of estimating $\alpha$ via the asymptotic growth rate of the $q$'th moment for low $q$ performs very well for $\alpha<1$, but has large errors for $\alpha>1$. This is due to the non-accuracy in determining the mean via the sample mean which is subtracted for $\alpha>1$. This undesirable property is alleviated when estimating the asymptotic growth rate of the twisted $q$'th moment, where $\alpha$ is well estimated for the whole range of $\alpha$. The standard $p$-variation performs well except near $\alpha=1$ and $\alpha=2$. It is more accurate than the twisted low moment method for $1<\alpha<1.5$. The modified $p$-variation has strong difficulties in estimating the anomalous diffusion near $\alpha=1$ and $\alpha>1.75$. We have tested that the bad performance of the modified $p$-variation method near $\alpha=1$ is due to the non-vanishing mean parameter $\mu=2$ and the asymmetry $\beta=1$. For $\mu=0$ and $\beta=0$ (and all other parameters unchanged), the modified $p$-variation method performs well near $\alpha=1$. The bad performance near the Brownian case $\alpha=2$ remains though for $\mu=0$ and $\beta=0$. As expected, the methods described in Section~\ref{sec.nolan} perform best in the case of noise-less {\em{i.i.d.}}\ random variables. The methods of quantile matching, linear regression of the empirical characteristic function and, in particular, the maximum likelihood estimator very accurately estimate $\alpha$ for the whole range of $\alpha$.\\

%

\begin{figure*}[!htbp]
\centering
\includegraphics[width = 0.475\columnwidth, height = 6cm]{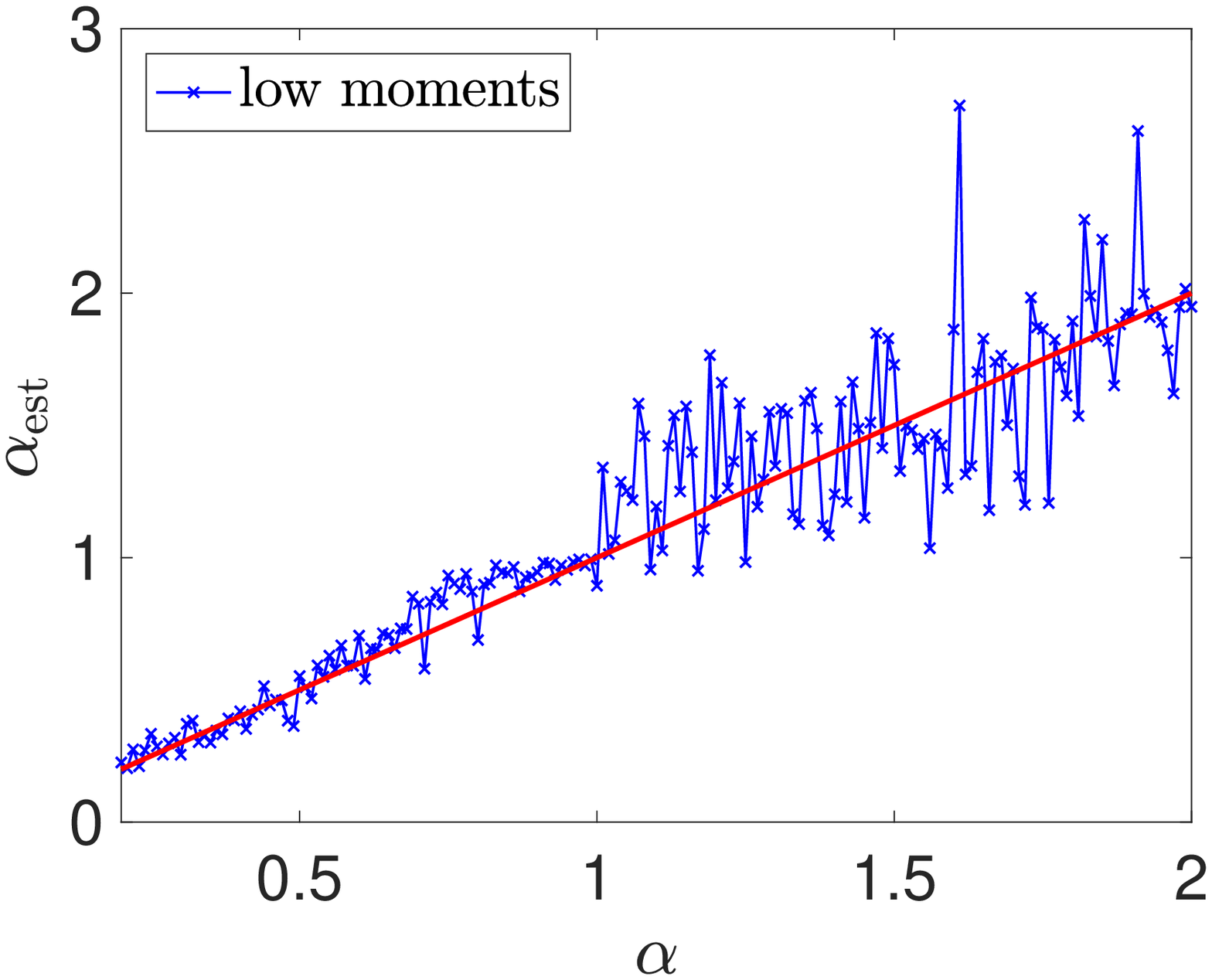}
\includegraphics[width = 0.475\columnwidth, height = 6cm]{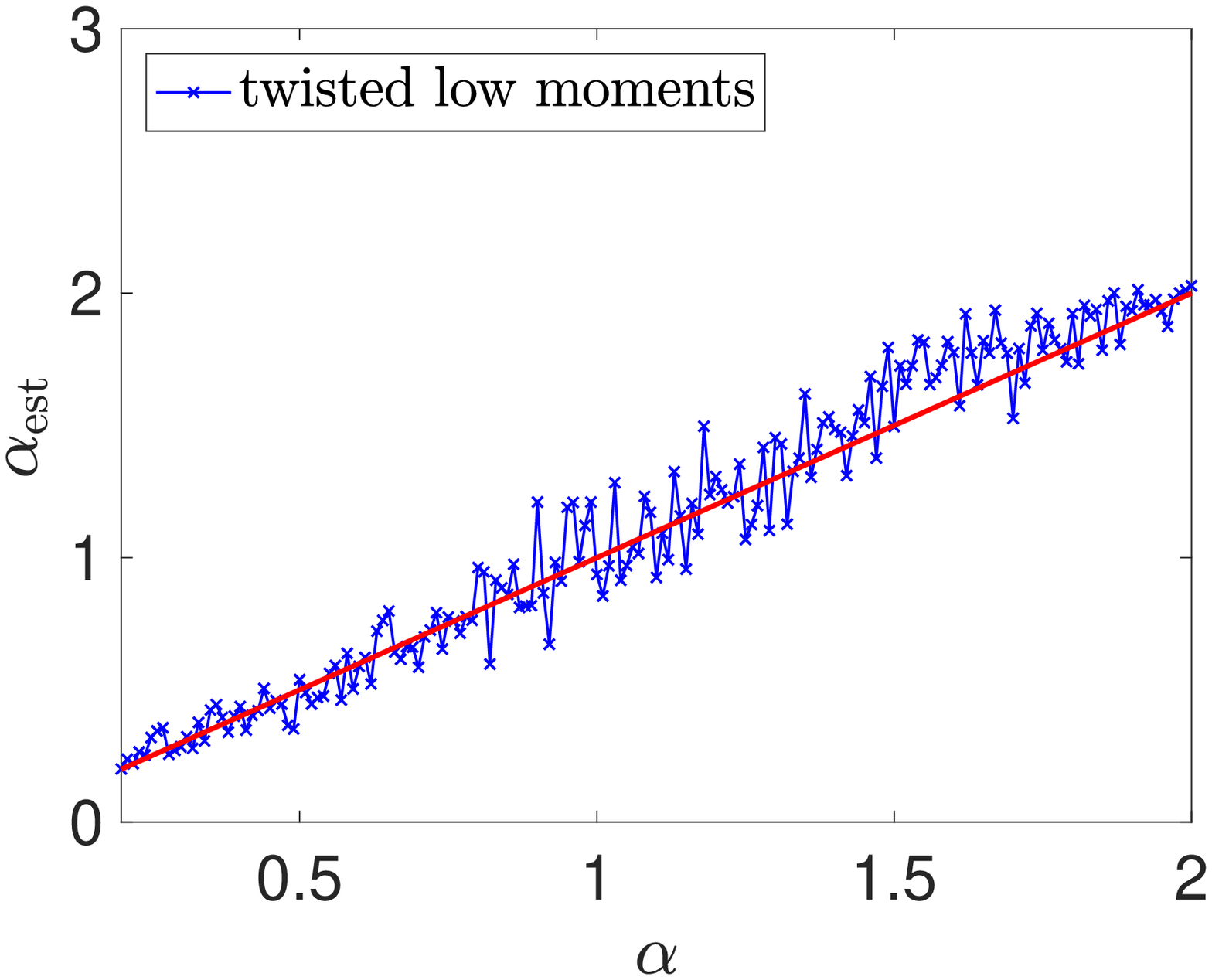}
\\
\includegraphics[width = 0.475\columnwidth, height = 6cm]{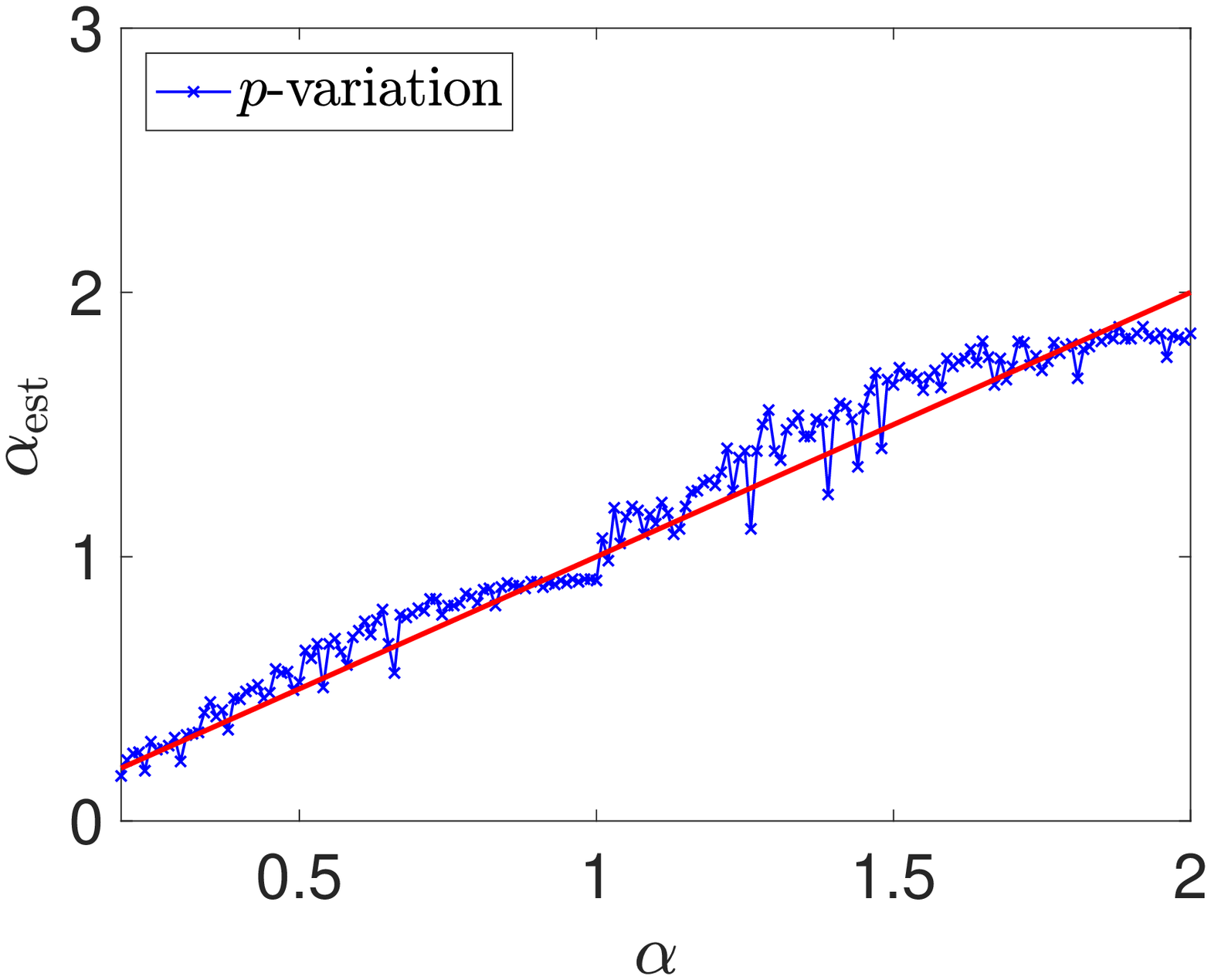}
\includegraphics[width = 0.475\columnwidth, height = 6cm]{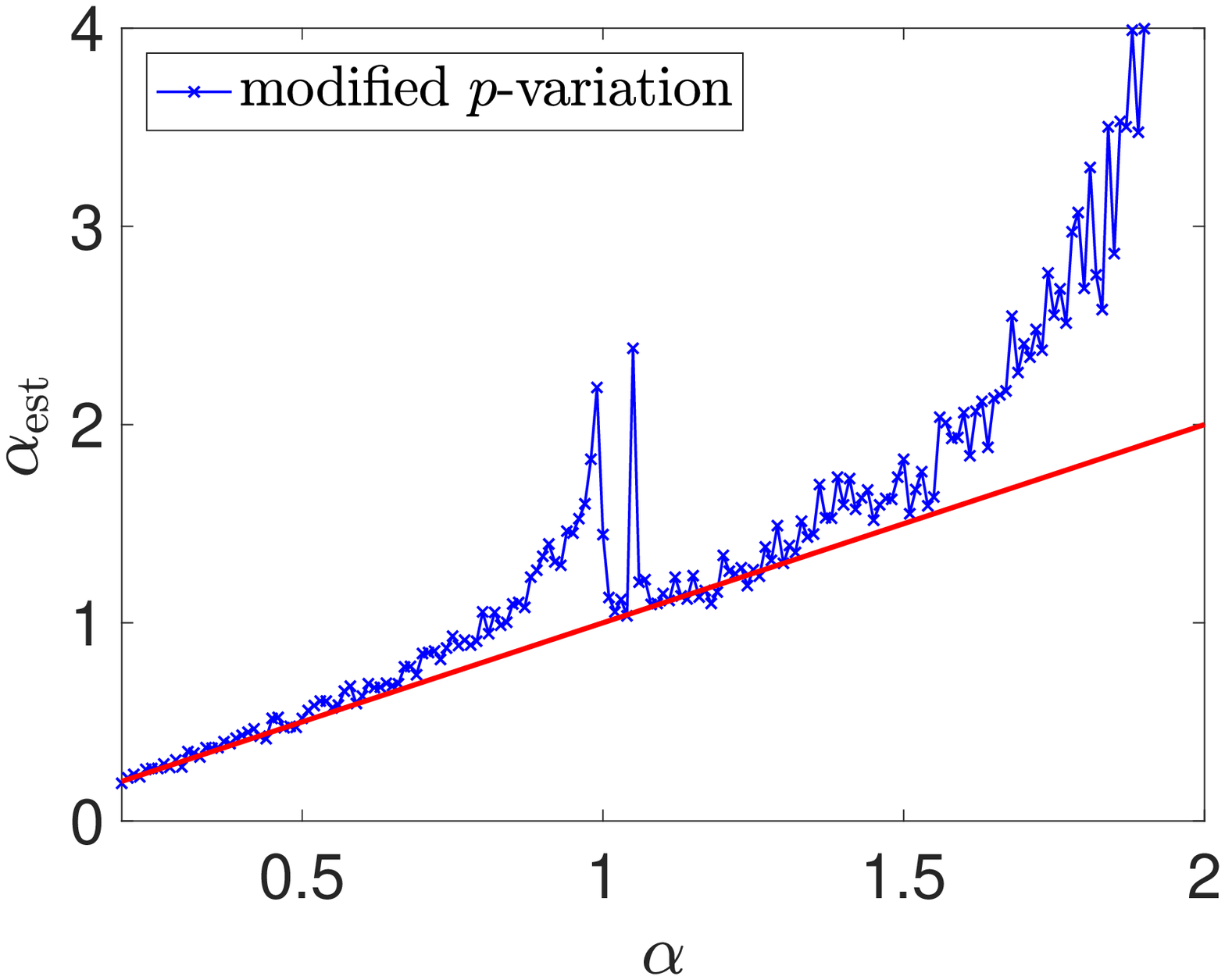}
\\
\includegraphics[width = 0.475\columnwidth, height = 6cm]{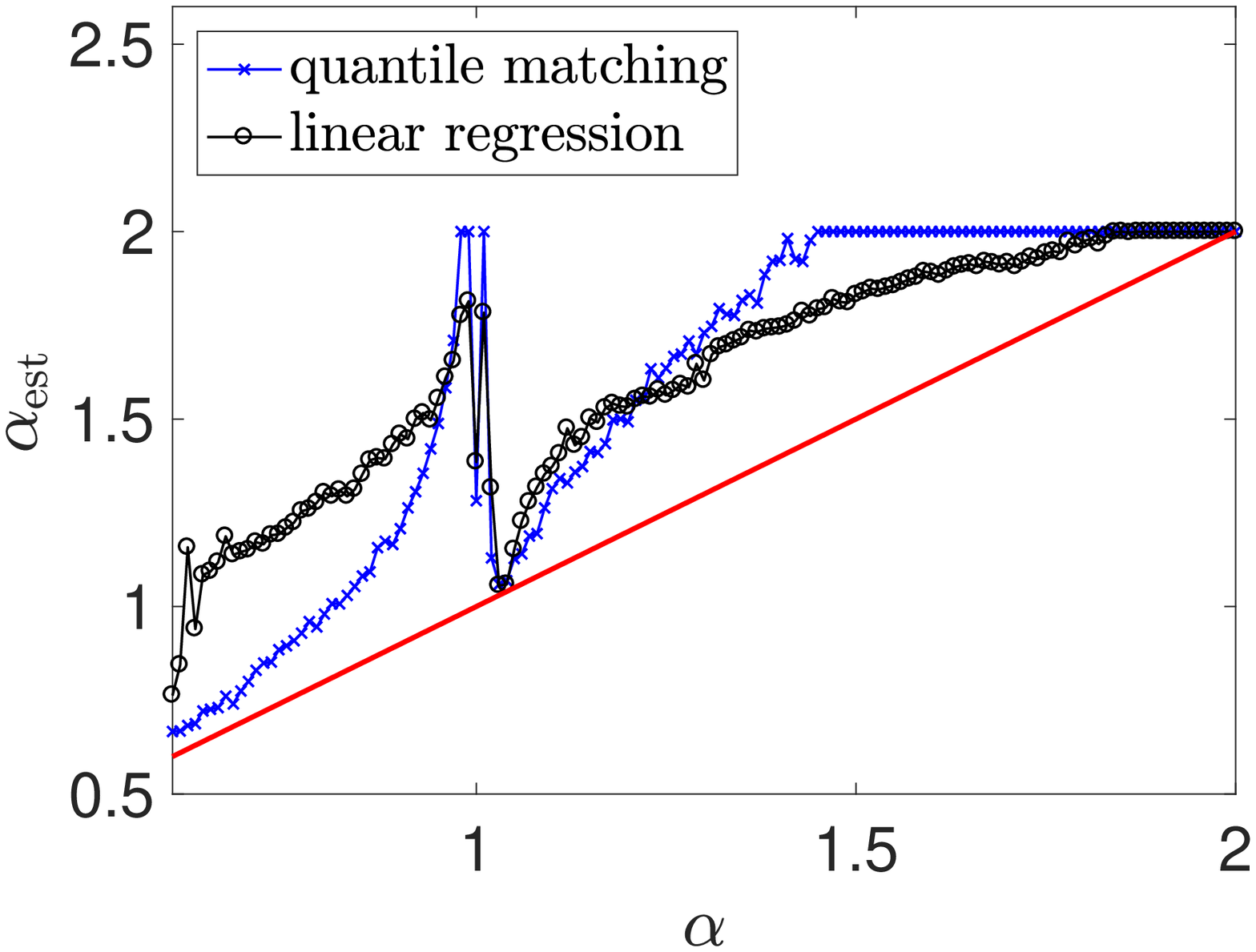}
\includegraphics[width = 0.475\columnwidth, height = 6cm]{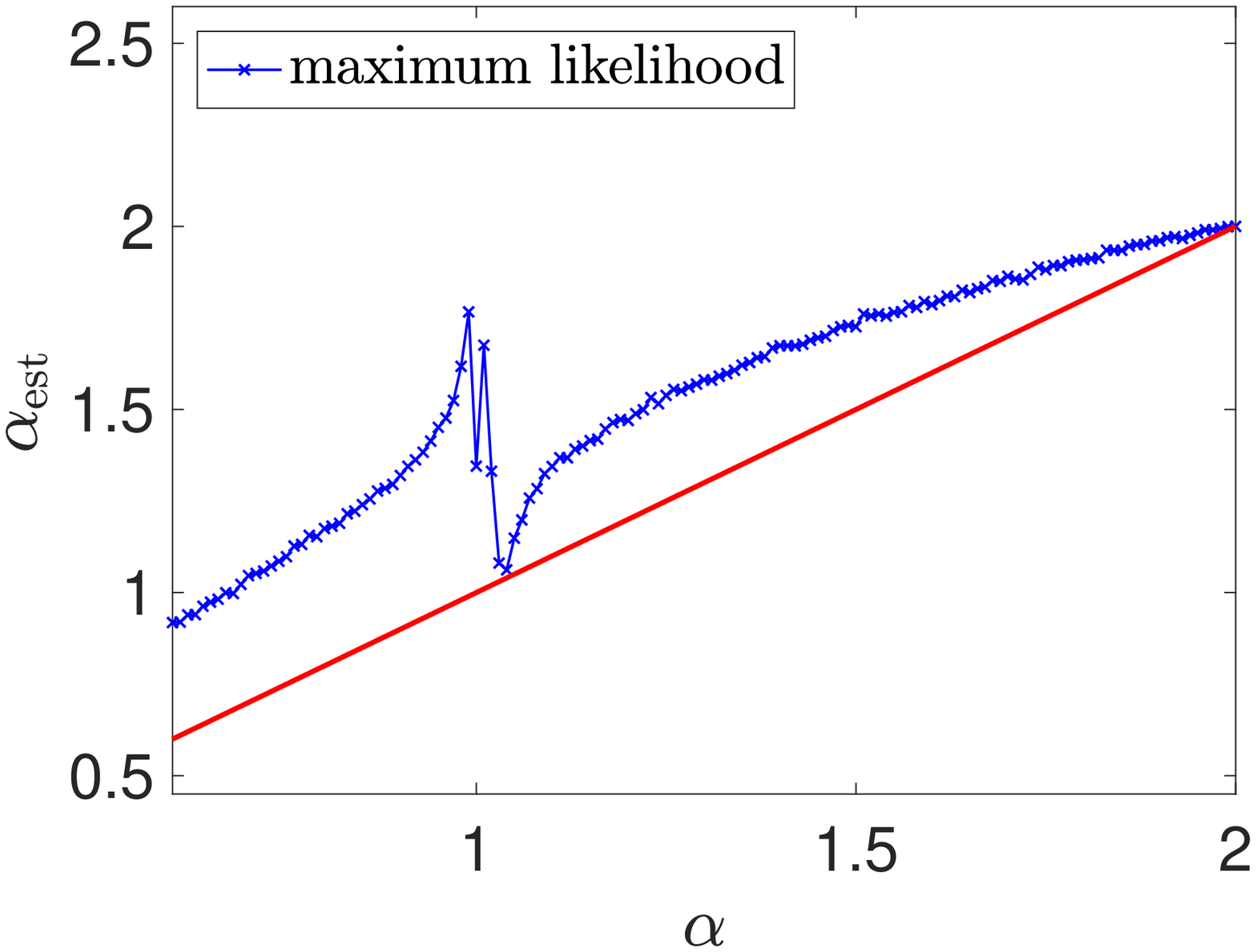}
\caption{Estimates $\alpha_{\rm{est}}$ as calculated for several values $\alpha\in(0.2,2)$ for {\em{i.i.d.}}\ observations with $50\%$ biased measurement noise. Top left: method of low $q$'th moment with $q=1/8$. Top right: method of twisted low $q$'th moment with $q=1/8$. Middle left: $p$-variation method. Middle right: modified $p$-variation method  (note the larger range on the $y$-axis). Bottom left: Methods using the empirical characteristic function. Bottom right: Maximum likelihood estimator. The straight lines show the theoretical answer $\alpha_{est}=\alpha$.}
\label{fig.IID_all_noise}
\end{figure*}

We also present results where we contaminate the observations $\varphi(n)$ by biased uniform noise according to $\varphi(n) \to (1+\eta u)\varphi(n)$ with $u\sim U(0,1)$ in Figure~\ref{fig.IID_all_noise}. We choose here relatively large additive measurement noise with $\eta=0.5$. The additive noise makes detection more difficult for determining the asymptotic growth rate of the $q$'th moment but much less so for the twisted low moment method. The standard $p$-variation also becomes less reliable for $1<\alpha<2$ when additive noise is included, consistent with the results reported in \cite{JeonEtAl13}. The modified $p$-variation exhibits some deteriorating sensitivity to additive noise. Contrary to the superior performance of the methods described in Section~\ref{sec.nolan} in the case of noise-less {\rm{i.i.d.}}\ random variables, these methods are not able to reliably estimate the stable parameter $\alpha$ when the data is contaminated by noise as shown in the bottom row of Figure~\ref{fig.IID_all_noise}.

%


\subsection{Results for the Pomeau-Manneville case}
\begin{figure*}[!htbp]
\centering
\includegraphics[width = 0.475\columnwidth, height = 6cm]{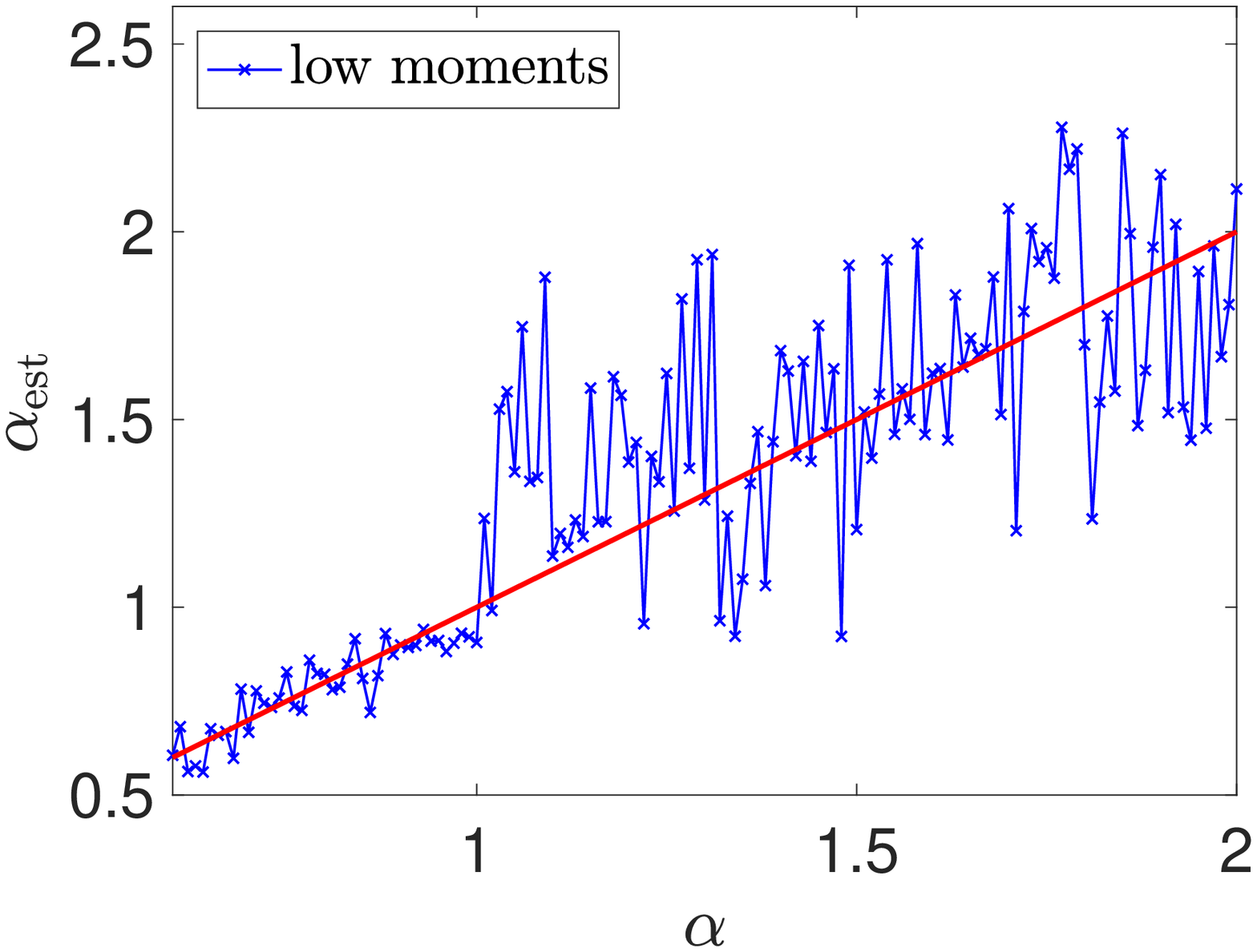}
\includegraphics[width = 0.475\columnwidth, height = 6cm]{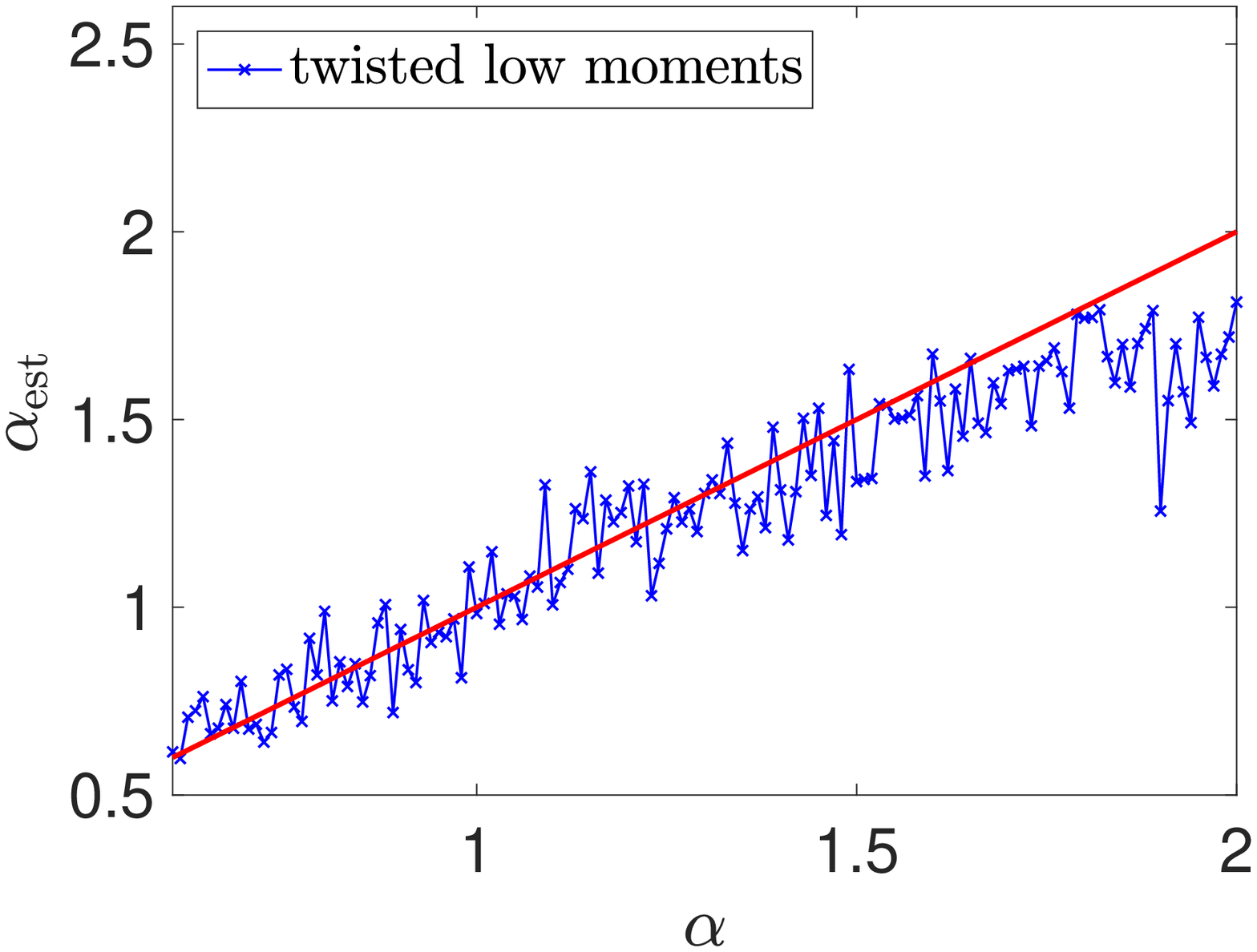}
\\
\includegraphics[width = 0.475\columnwidth, height = 6cm]{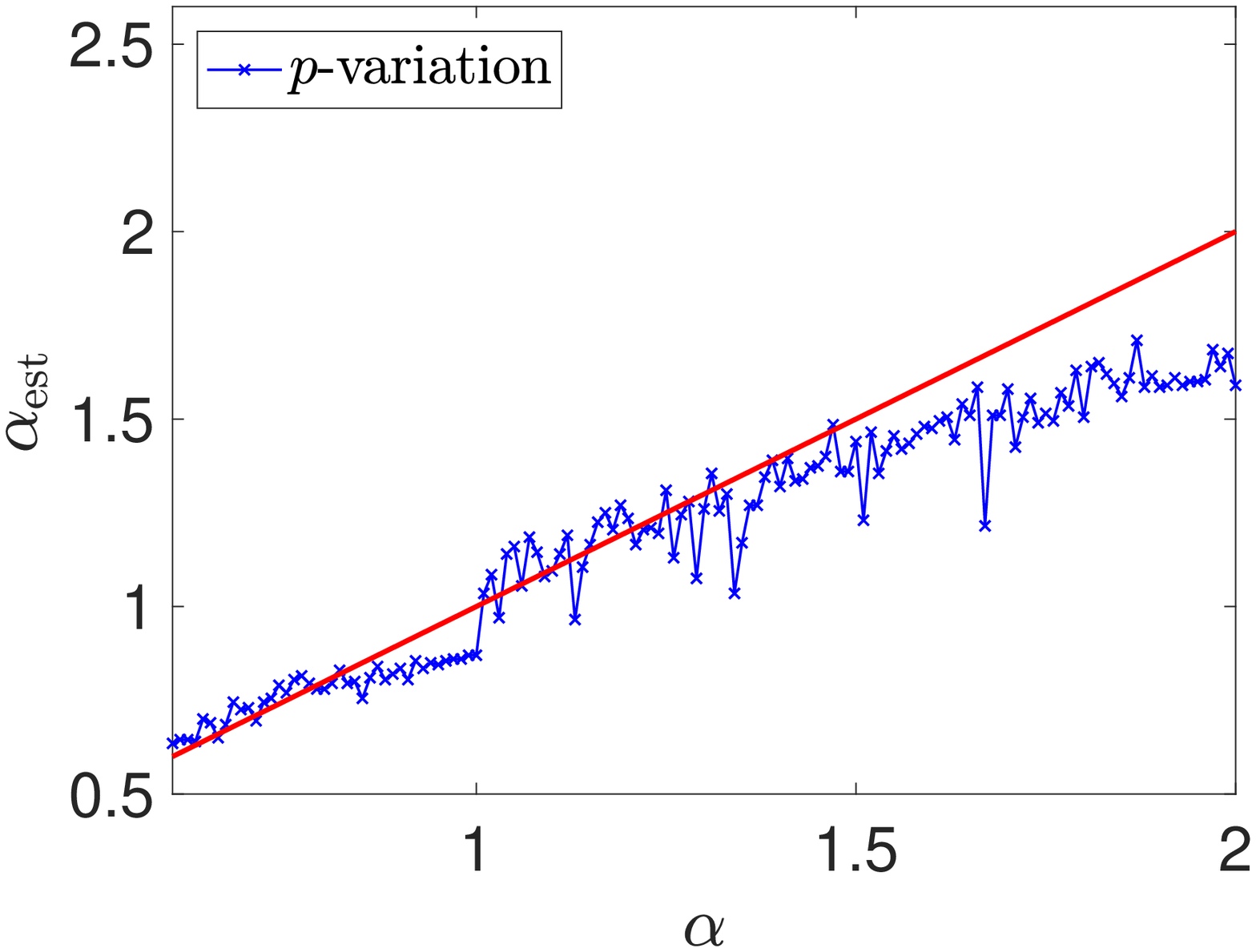}
\includegraphics[width = 0.475\columnwidth, height = 6cm]{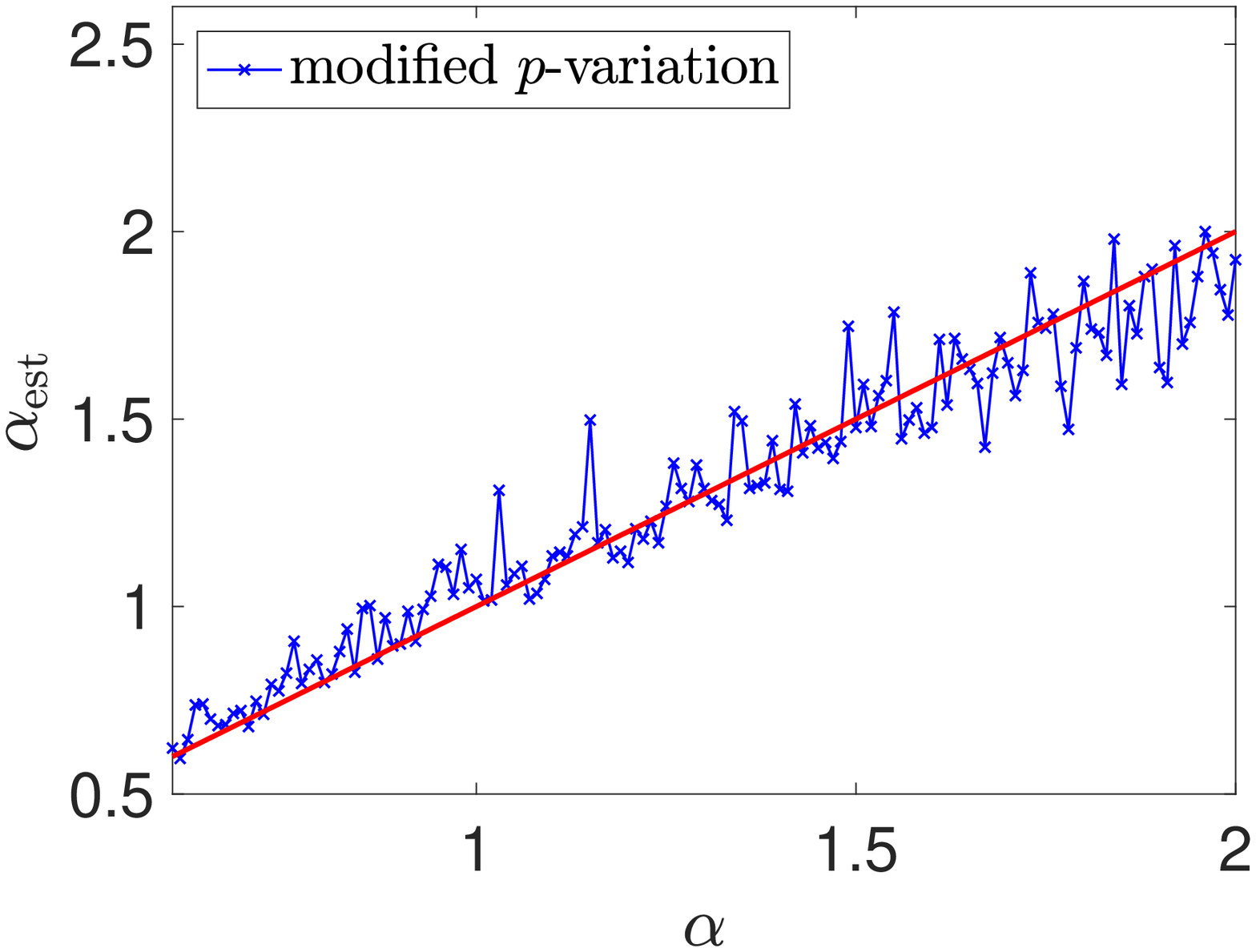}
\\
\includegraphics[width = 0.475\columnwidth, height = 6cm]{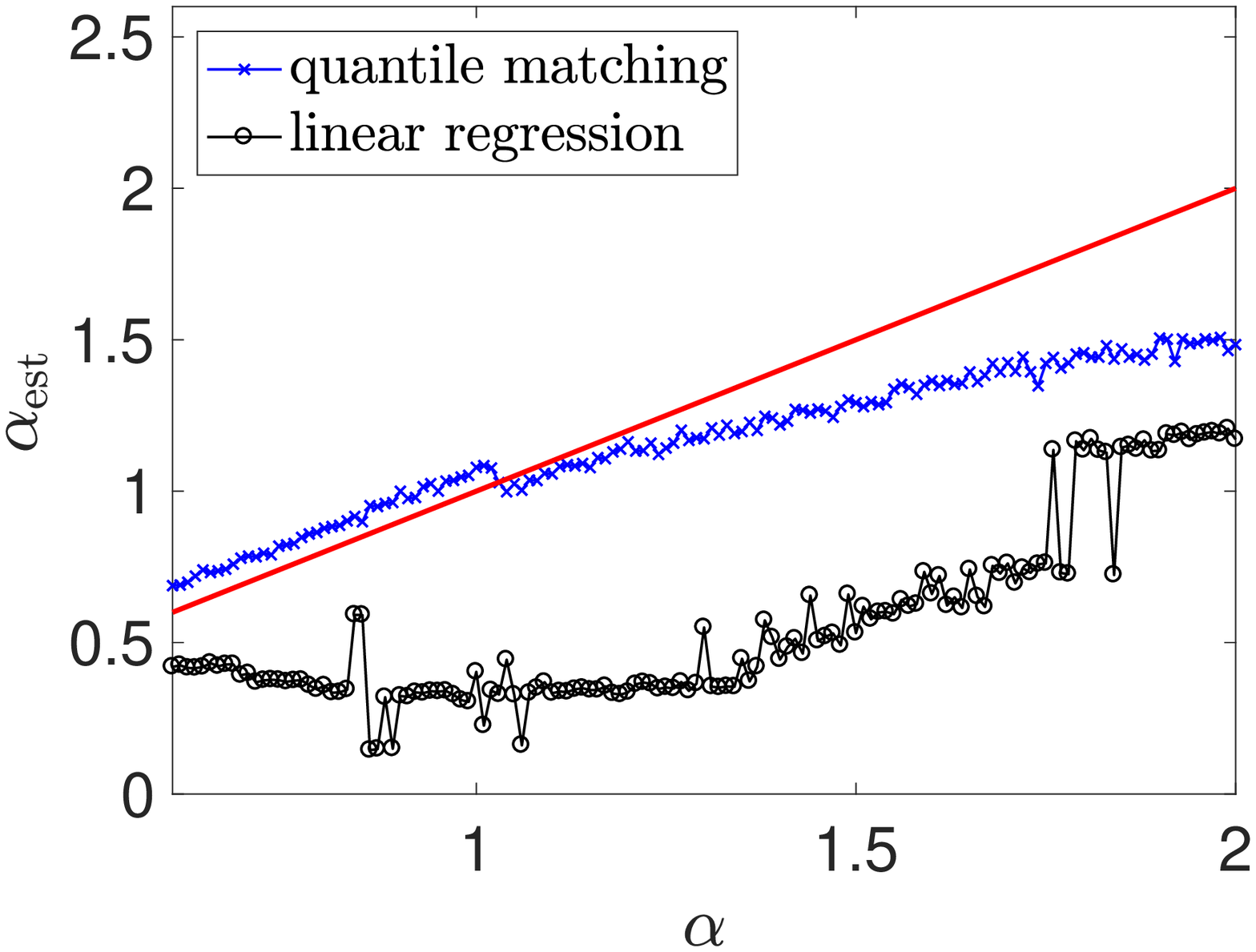}
\includegraphics[width = 0.475\columnwidth, height = 6cm]{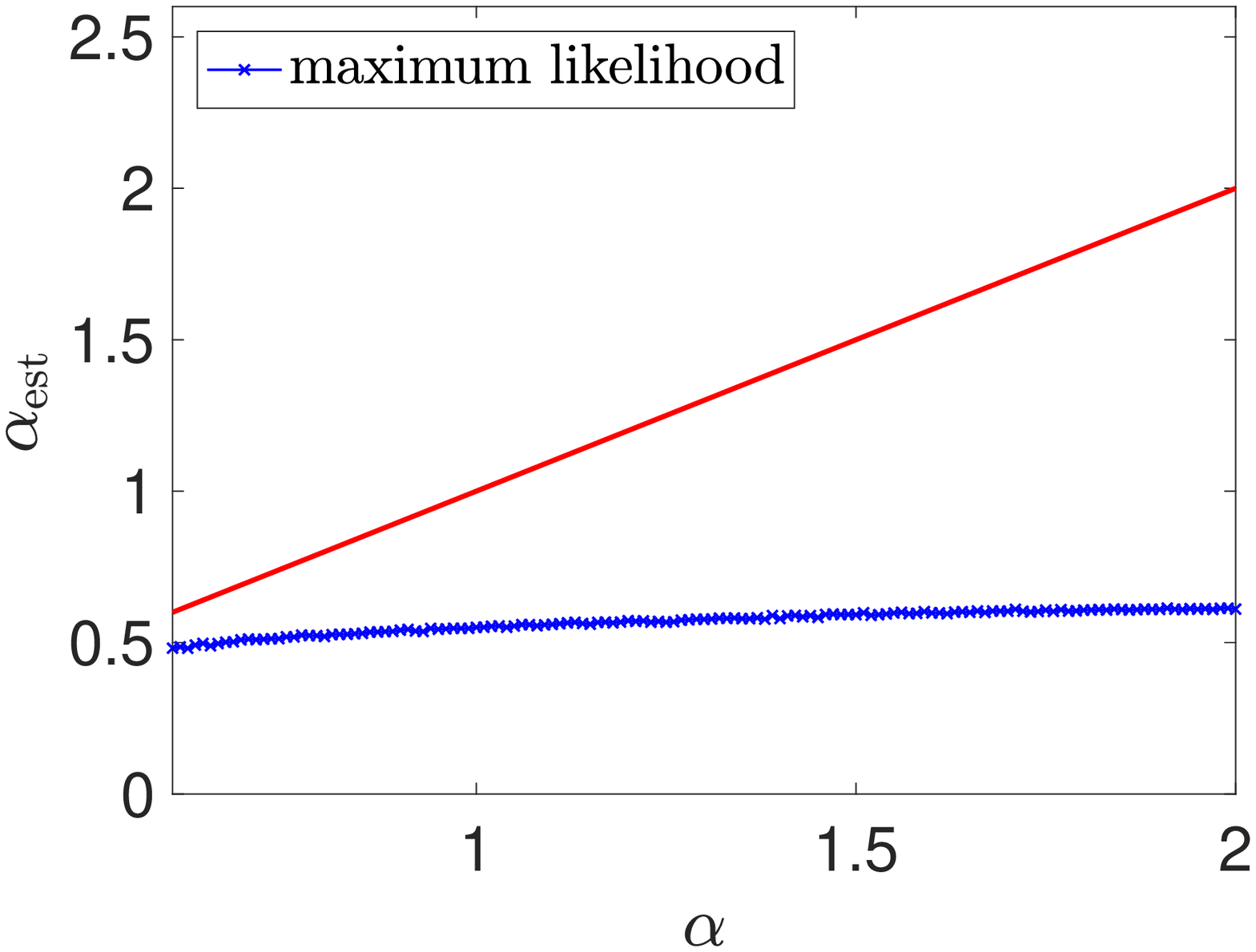}
\caption{Estimates $\alpha_{\rm{est}}$ as calculated for several values $\alpha\in(0.6,2)$ for the Pomeau-Manneville map (\ref{e.PM}). Top left: method of low $q$'th moment with $q=1/8$. Top right: method of twisted low $q$'th moment with $q=1/8$. Middle left: $p$-variation method. Middle right: modified $p$-variation method. Bottom left: Methods using the empirical characteristic function. Bottom right: Maximum likelihood estimator. The straight lines show the theoretical answer $\alpha_{est}=\alpha$.}
\label{fig.PM_all}
\end{figure*}

We use a time series of length $N=25,000$ constructed from the Pomeau-Manneville map (\ref{e.PM}) for $\alpha \in (0.6,2)$. We choose $\varphi(y)=1+y$. Figure~\ref{fig.PM_all} shows the analogous results to Figure~\ref{fig.IID_all}.

In the deterministic case, we observe the same behaviour of low moments as in the {\em{i.i.d.}}\ case where anomalous diffusion is very well classified for $\alpha<1$ but not so well for $\alpha>1$ where the error in estimating the sample mean has a detrimental effect. The twisted low moment method performs well, except near the Brownian case of $\alpha=2$ where it underestimates the anomalous scaling coefficient. The slow convergence may be related to cross-correlation effects that arise in the diffusion parameter via the Green-Kubo formula. Such cross-correlations are not present in the superdiffusive case $\alpha<2$ \cite{Gouezel04}. The $p$-variation also does not perform well. Near $\alpha=1$ and the Brownian case $\alpha=2$ the $p$-variation strongly flattens and underestimates the anomalous scaling coefficient. The modified $p$-variation, in contrast, performs well for the whole range of $\alpha$. As in the case of noisy {\em{i.i.d.}}\ variables, the standard {\em{i.i.d.}}\ estimation methods described in Section~\ref{sec.nolan} do not reliably estimate the stable parameter $\alpha$ for the whole range of $\alpha$. Curiously, the quantile matching method performs well for $\alpha<1$.\\

%

\begin{figure*}[!htbp]
\centering
\includegraphics[width = 0.475\columnwidth, height = 6cm]{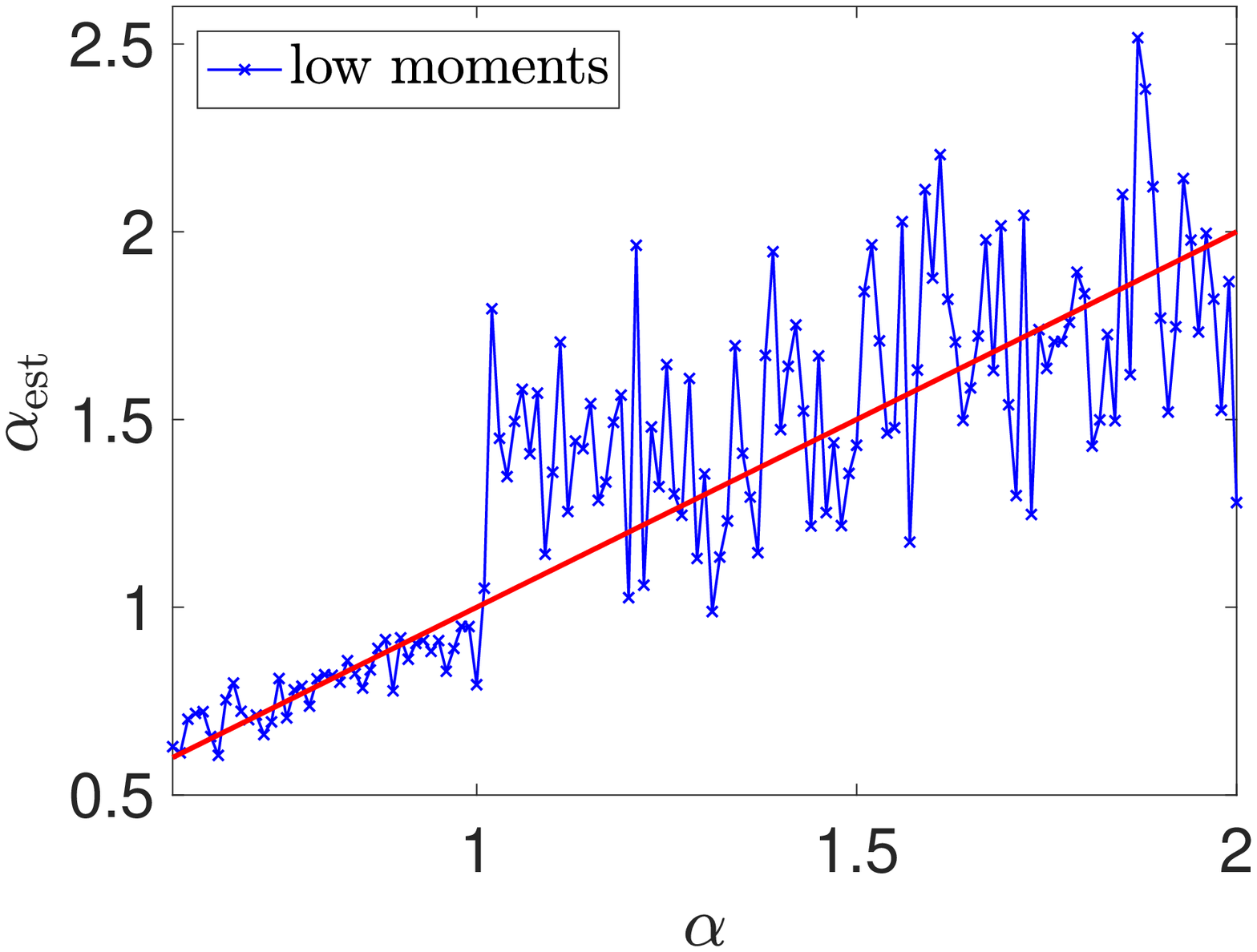}
\includegraphics[width = 0.475\columnwidth, height = 6cm]{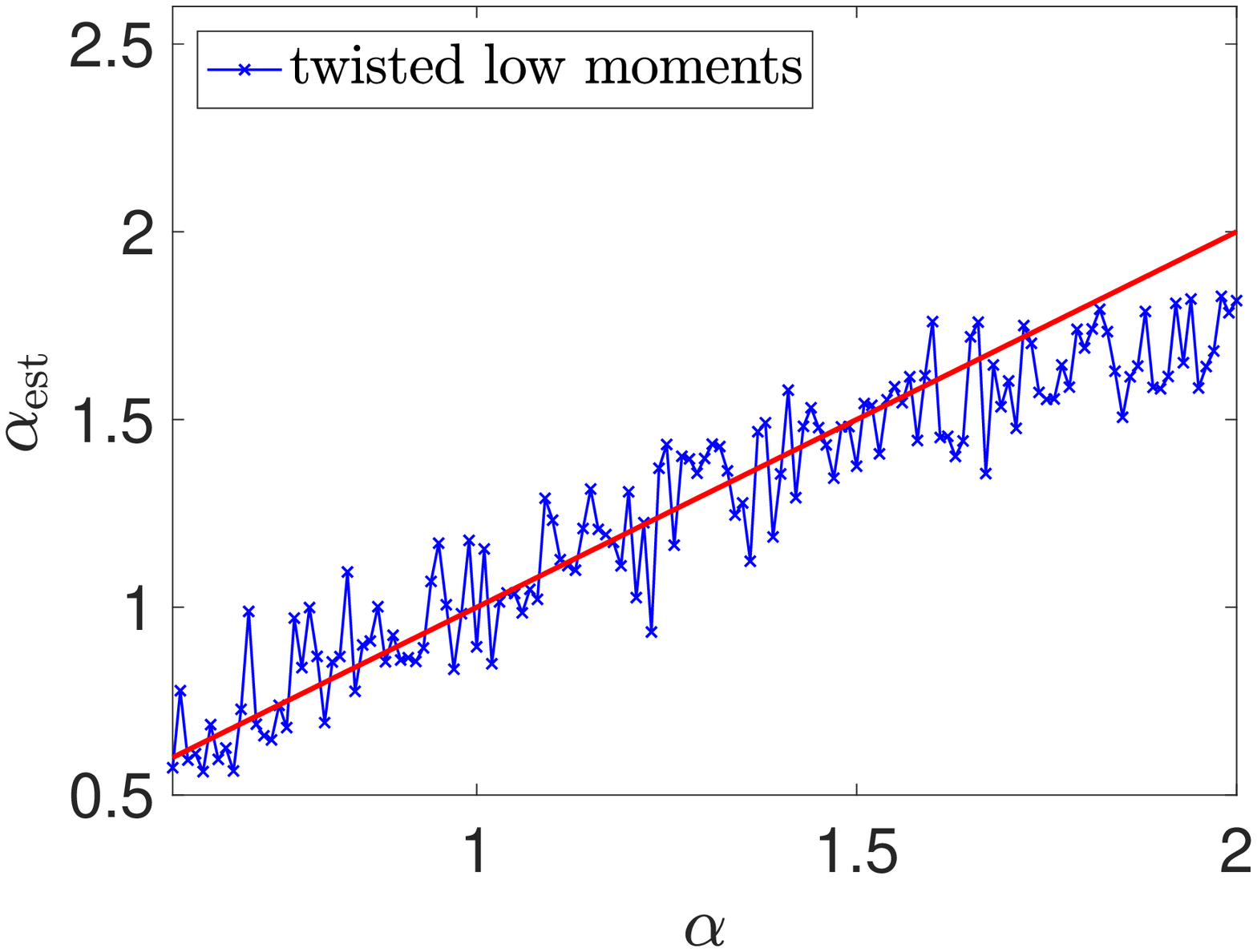}
\\
\includegraphics[width = 0.475\columnwidth, height = 6cm]{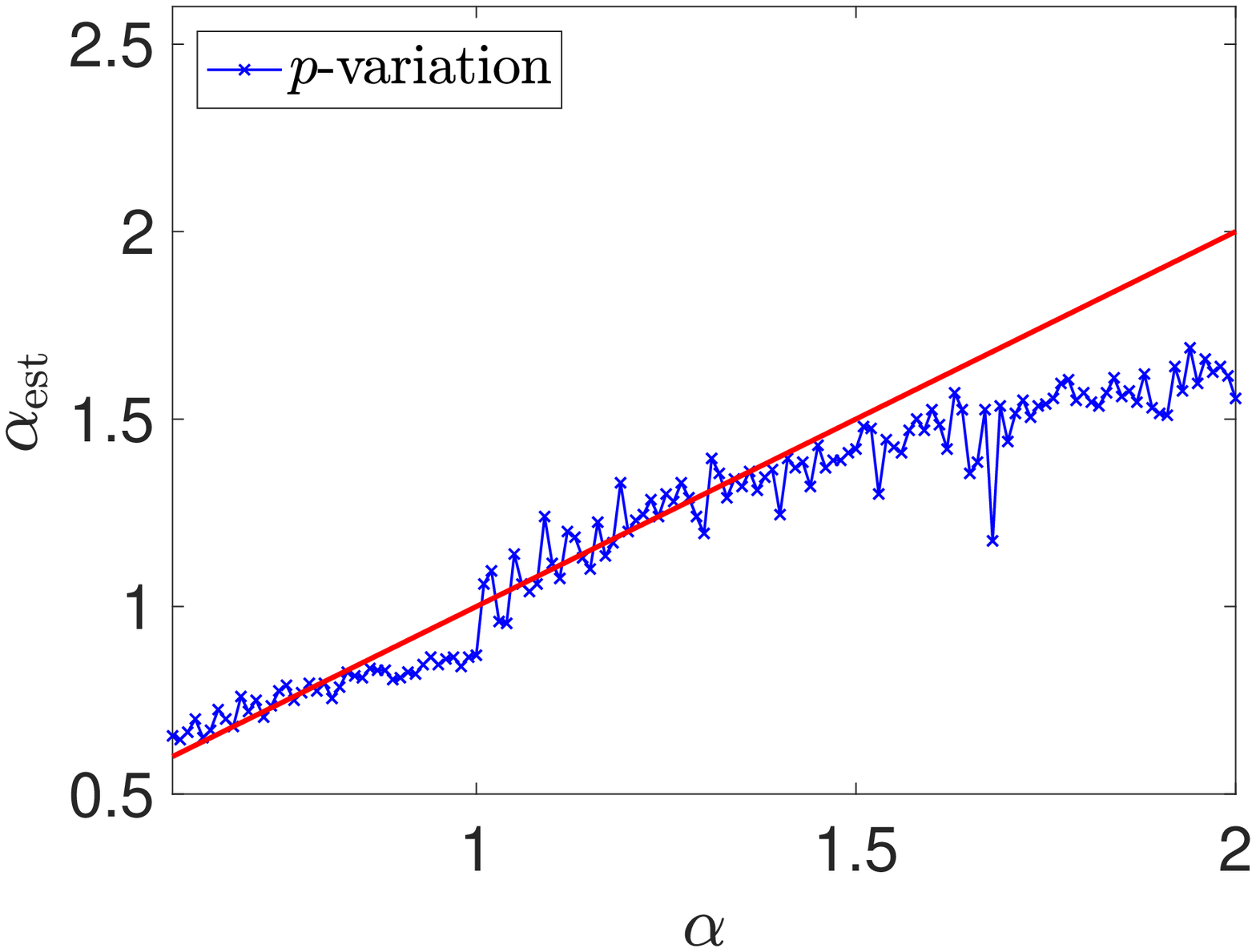}
\includegraphics[width = 0.475\columnwidth, height = 6cm]{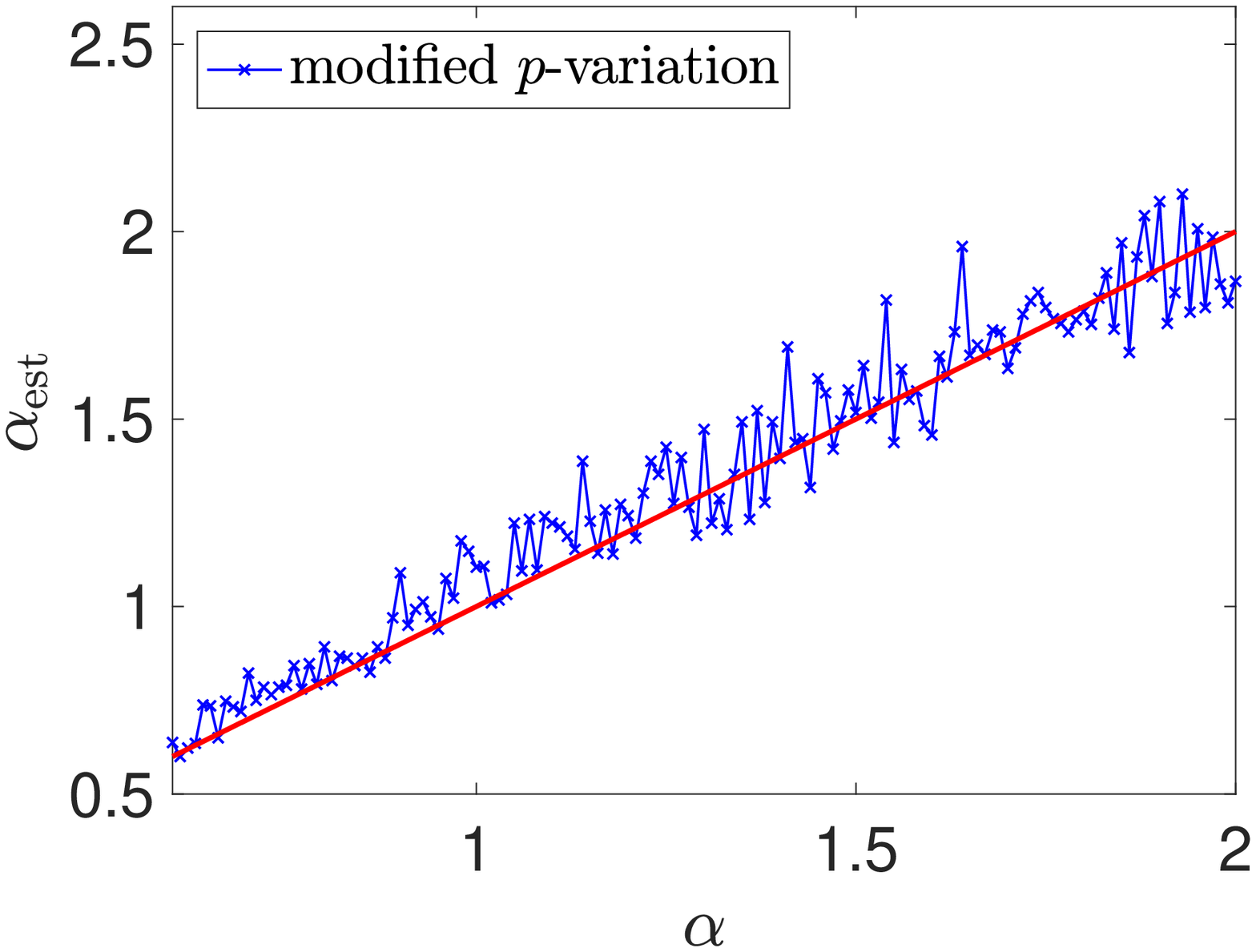}
\\
\includegraphics[width = 0.475\columnwidth, height = 6cm]{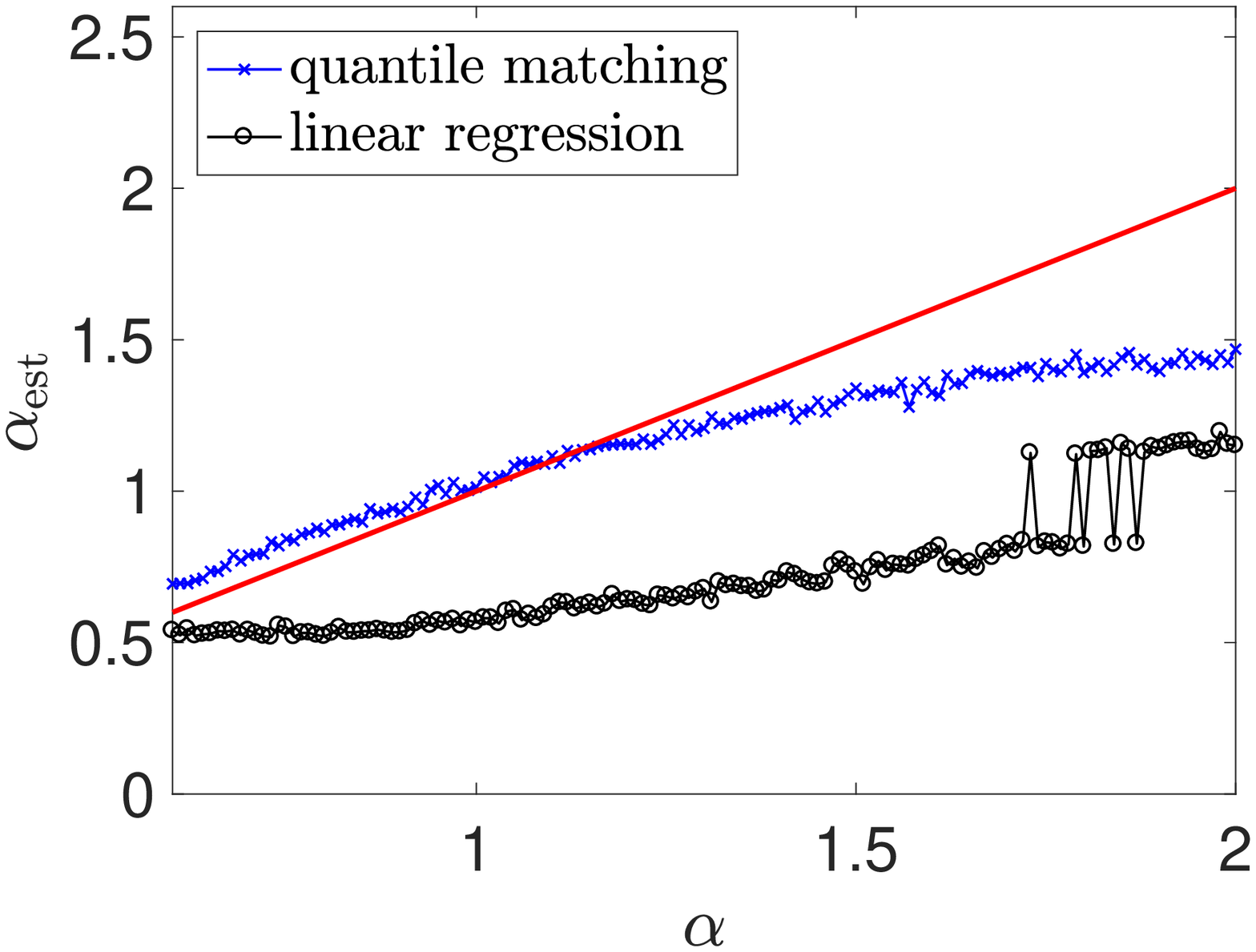}
\includegraphics[width = 0.475\columnwidth, height = 6cm]{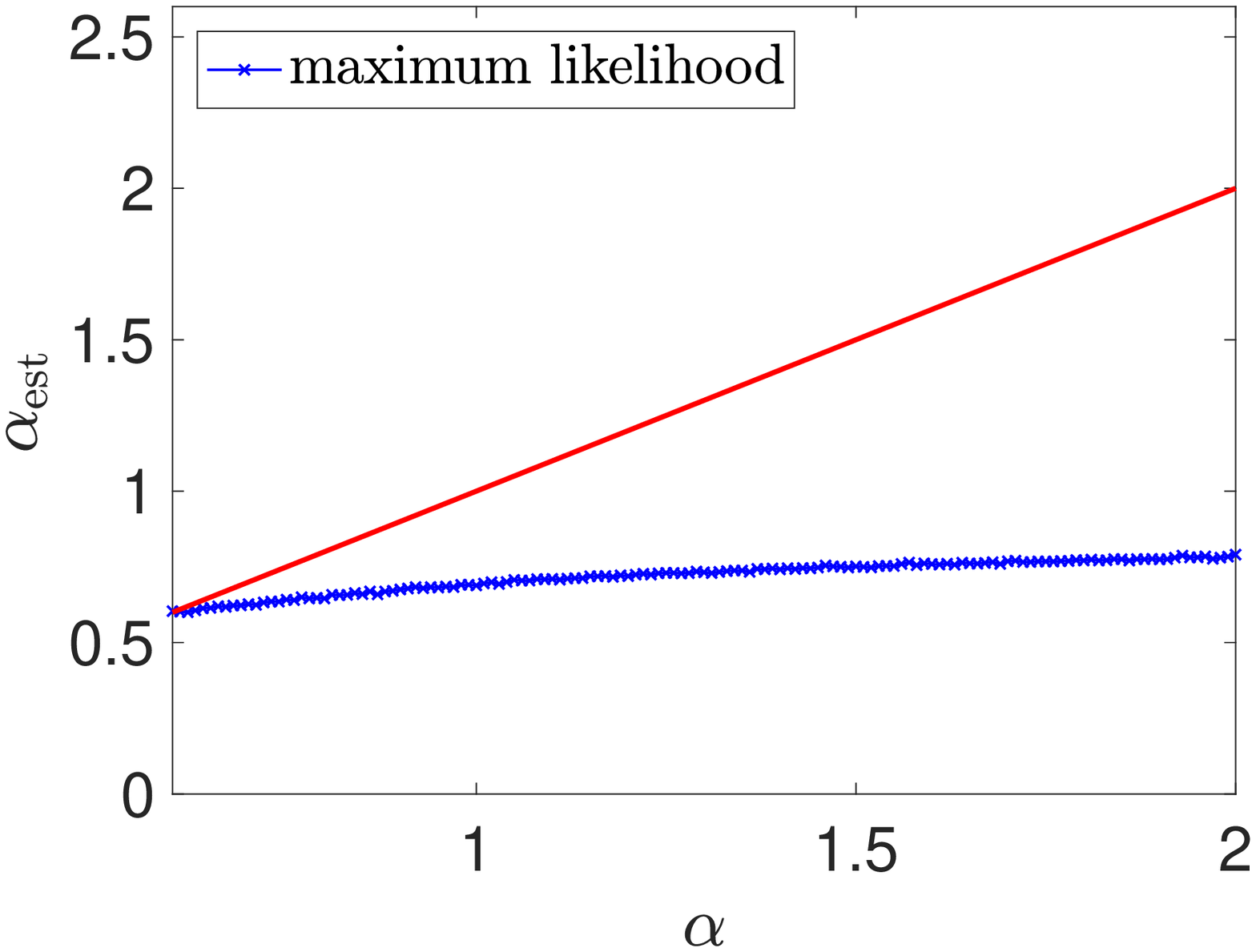}
\caption{Estimates $\alpha_{\rm{est}}$ as calculated for several values $\alpha\in(0.6,2)$ for the Pomeau-Manneville map (\ref{e.PM}) with $50\%$ biased measurement noise. Top left: method of low $q$'th moment with $q=1/8$. Top right: method of twisted low $q$'th moment with $q=1/8$. Middle left: $p$-variation method. Middle right: modified $p$-variation method. Bottom left: Methods using the empirical characteristic function. Bottom right: Maximum likelihood estimator. The straight lines show the theoretical answer $\alpha_{est}=\alpha$.}
\label{fig.PM_all_noise}
\end{figure*}

Again, we also present results for observations which have been contaminated with biased uniform noise with $\eta=0.5$ in Figure~\ref{fig.PM_all_noise}. As in the {\em{i.i.d.}}\ case, the performance of the low moment method and the $p$-variation method is diminished by  the additive measurement noise. The performance of the twisted low moment method and the modified $p$-variation method, however, are robust against additive measurement noise. The standard {\em{i.i.d.}}\ estimation methods described in Section~\ref{sec.nolan} fail to reliably estimate the stable parameter $\alpha$ for the whole range of $\alpha$. Again, the quantile matching method performs well for $\alpha<1$. The method of linear regression of the empirical characteristic function and the maximum likelihood method significantly underestimate the value of $\alpha$.\\

%


\section{Summary and Discussion}
\label{sec.sum}

We have introduced a new method to quantitatively estimate the degree of anomalous superdiffusion. Our method uses the asymptotic growth rate of a twisted low moment derived from the data rotated with a periodic deterministic signal.\\ We established that the standard method of estimating the growth-rate of the mean-square displacement is not able to reliably distinguish superdiffusion from normal diffusion in finite time series. We have compared our method then with ive other methods, a method based on (untwisted) low moments, two versions of the $p$-variation method as well as the standard estimators analysing the empirical characteristic function and estimating the maximum likelihood developed for {\em{i.i.d.}}\ random variables.\\ Whereas the standard methods such as quantile matching, linear regression of the empirical characteristic function and maximum likelihood estimators are by far superior in estimating the stable parameter $\alpha$ in the case of noise-free {\em{i.i.d.}}\ random variables, they fail in the case of noisy {\em{i.i.d.}} random variables and/or deterministically generated variables. Our numerical simulations on noisy {\em{i.i.d.}}\ data and data generated deterministically from weakly chaotic Pomeau-Mannneville maps, reveal that our new method and the modified $p$-variation as proposed in \cite{HeinImkellerPavlyukevich} perform best and are most robust to additive measurement noise, which is inevitable in any real-world application. The modified $p$-variation and our newly proposed twisted low moment method have been shown to have complementary advantages. Whereas the modified $p$-variation performs very well in the case of deterministic data, it did less so for the {\em{i.i.d.}}\ case, in particular for values of $\alpha$ near $1$ and $2$. In contrast, our new method performs well in the case of {\em{i.i.d.}}\ random variables, but becomes less accurate in the deterministic case for values of $\alpha$ approaching Brownian diffusion with $\alpha=2$. We therefore propose our method to be used in conjunction with the $p$-variation to gain further insights into the quantitative analysis of anomalous diffusion from time series.\\ The computational cost involved in applying those methods varies significantly. 
The standard $p$-variation method and the method of estimating the asymptotic growth rate of a low moment are the least computationally demanding methods. For the twisted low moments, one needs to cycle over typically $100$ different values of the frequency $c$ of the periodic signal. The modified $p$-variation method requires cycling through values of $\sigma$, which requires tuning over a large range of values. Despite the variation in the computational cost of the methods, all the methods use only a single sample time series.   



\paragraph{Acknowledgements}
We would like to thank John Nolan for generously sharing his software package {\rm{STABLE}} with us. 
This research was supported in part by an International Research Collaboration Award at the University of Sydney. 
The research of IM was supported in part by a European Advanced Grant StochExtHomog (ERC AdG 320977).


\appendix 


\setcounter{section}{0}

\section{Modified $p$-variation}
\label{sec-appb}
We recall here Theorem 2.1 from \cite{HeinImkellerPavlyukevich}
\begin{thm}
For an $\alpha$-stable process $X_t$ with $X_t \sim S_\alpha(\beta,0,\sigma)$, we have for $p>\alpha/2$ that its $p$-variation $V_p^n(t)$ converges in the Skorohod topology with
\begin{align*}
V_p^n (t) - ntB_n(\alpha,p)  \rightarrow_{d} X^\prime_t  \ \mathrm{as} \ n \rightarrow  \infty ,
\end{align*}
where $X^\prime_t\sim S_{\frac{\alpha}{p}}(1,0,\sigma^\prime)$ with spread parameter
\begin{align*}
\sigma^\prime = \begin{cases} \sigma^p \left( \frac{\cos(\frac{\pi\alpha}{2p})\Gamma(1-\frac{\alpha}{p})}{\cos(\frac{\pi\alpha}{2p})\Gamma(1-\alpha)} \right)^{p/\alpha} & p\not=\alpha \\ 
\sigma & p=\alpha \end{cases}
,
\end{align*}
and normalising sequence 
\begin{align*}\label{bn}
B_n(\alpha, p) = \begin{cases} n^{-p/\alpha}{\bf E} |X|^p & p \in (\alpha/2, \alpha) \\ 
{\bf E}\sin\left(n^{-1}|X|^\alpha \right) & p=\alpha \\
0 & p>\alpha 
\end{cases}
.
\end{align*}
\end{thm}


\section*{References}

\providecommand{\newblock}{}

\end{document}